\documentclass[twocolumn,tighten]{aastex6}
\shorttitle{A $\gamma$-ray loud eclipsing X-ray binary}
\shortauthors{Strader \etal~}
\def\etal{{et al.}}

\def\arcsec{\char'175 }

\def\hub{\ifmmode H_\circ\else H$_\circ$\fi}
\usepackage{subfigure}
\usepackage{amsmath}
\usepackage{graphics}
\usepackage{hyperref}
\usepackage{breakurl} 

\def\ltsima{$\; \buildrel < \over \sim \;$}
\def\simlt{\lower.5ex\hbox{\ltsima}} 
\def\gtsima{$\; \buildrel > \over \sim \;$}
\def\simgt{\lower.5ex\hbox{\gtsima}} 
\def\arcsec{\hbox{$^{\prime\prime}$}}

\newcommand{\flux}{\,erg\,s$^{-1}$\,cm$^{-2}$}

\newcommand{\cm}{\,cm$^{-2}$}
\newcommand{\nh}{$N_\mathrm{H}$}

\begin{document}

\title{
A new $\gamma$-ray loud, eclipsing low-mass X-ray binary}

\author{
Jay Strader\altaffilmark{1},
Kwan-Lok Li\altaffilmark{1},
Laura Chomiuk\altaffilmark{1}, 
Craig O.~Heinke\altaffilmark{2},
Andrzej Udalski\altaffilmark{3},
Mark Peacock\altaffilmark{1},
Laura Shishkovsky\altaffilmark{1},
Evangelia Tremou\altaffilmark{1}
}
\altaffiltext{1}{Department of Physics and Astronomy, Michigan State University, East Lansing, MI 48824, USA; strader@pa.msu.edu}
\altaffiltext{2}{Department of Physics, CCIS 4-183, University of Alberta, Edmonton, AB T6G 2E1, Canada}
\altaffiltext{3}{Warsaw University Observatory, Al.~Ujazdowskie 4, 00-478 Warszawa, Poland}

\begin{abstract}

We report the discovery of an eclipsing low-mass X-ray binary at the center of the 3FGL error ellipse of the unassociated \emph{Fermi}/Large Area Telescope $\gamma$-ray source 3FGL J0427.9--6704. Photometry from OGLE and the SMARTS 1.3-m telescope and spectroscopy from the SOAR telescope have allowed us to classify the system as an eclipsing low-mass X-ray binary (P = 8.8 hr) with a main sequence donor and a neutron star accretor. Broad double-peaked H and He emission lines suggest the ongoing presence of an accretion disk. Remarkably, the system shows shows separate sets of absorption lines associated with the accretion disk and the secondary, and we use their radial velocities to find evidence for a massive ($\sim 1.8$--1.9 $M_{\odot}$) neutron star primary. In addition to a total X-ray eclipse of duration

 $\sim 2200$ s observed with \emph{NuSTAR}, the X-ray light curve also shows properties similar to those observed among known transitional millisecond pulsars: short-term variability, a hard power-law spectrum ($\Gamma \sim 1.7$), and a comparable 0.5--10 keV luminosity ($\sim 2.4 \times 10^{33}$ erg s$^{-1}$). We find tentative evidence for a partial ($\sim 60\%$) $\gamma$-ray eclipse at the same phase as the X-ray eclipse, suggesting the $\gamma$-ray emission may not be confined to the immediate region of the compact object. The favorable inclination of this binary is promising for future efforts to determine the origin of $\gamma$-rays among accreting neutron stars.

\end{abstract}
 

\section{Introduction}

For 8 years the \emph{Fermi} Large Area Telescope (LAT) has surveyed the sky in GeV $\gamma$-rays. Among its most exciting discoveries is that compact binaries containing millisecond pulsars are ubiquitous $\gamma$-ray emitters (Abdo et al.~2013). In most systems this emission appears to be solely associated with the pulsar. The $\gamma$-ray emission serves as a beacon for the discovery of pulsar binaries typically missed by radio surveys: ``redback" or ``black widow"  systems where the pulsar is ablating a low (or very-low) mass hydrogen-rich companion (Roberts 2013). The excess material in these systems hinders detection as radio pulsars but does not affect the $\gamma$-ray emission.

Three redbacks belong to a yet smaller subclass of systems known as transitional millisecond pulsars, all of which have been observed to switch between rotational-powered states with radio-detected millisecond pulsars and accretion-powered states with an accretion disk and no observed radio pulsar emission (Archibald et al.~2009; Papitto et al.~2013; Bassa et al.~2014). This represents not only a confirmation of the general theory for the recycling of millisecond pulsars, but an opportunity to study the accretion and recycling process in real time.

Among the intriguing phenomenology of these binaries is that, in two of the three systems, the $\gamma$-ray emission becomes \emph{brighter} when the neutron star is accreting in the disk state (Stappers et al.~2014; Takata et al.~2014; Johnson et al.~2015), strongly implying that it does not originate in the pulsar magnetosphere as for rotationally-powered pulsars. The source of the $\gamma$-ray emission in these systems is not yet known.

 \emph{Fermi}-LAT $\gamma$-ray emission has also led to the identification of two binaries that show similarities to the transitional millisecond pulsars: 1RXS J154439.4--112820 is a low-mass X-ray binary with X-ray and optical properties that resemble the confirmed transitional systems (Bogdanov \& Halpern 2015), while 1FGL J1417.7--4407 contains a radio pulsar but also shows evidence for an accretion disk (Strader et al.~2015; Camilo et al.~2016). It is clear that follow-up of unidentified \emph{Fermi}-LAT $\gamma$-ray sources continues to be a promising route to find interesting compact binaries. 
 
This paper reports the discovery and characterization of a low-mass X-ray binary associated with the \emph{Fermi} source 3FGL J0427.9--6704. We show that this system not only bears a remarkable likeness to the known transitional millisecond pulsars, but also exhibits optical, X-ray, and possibly $\gamma$-ray eclipses, allowing novel constraints on the properties of the binary and the origin of the $\gamma$-ray emission.

\newpage
\section{Data}

\begin{figure*}[t]
\includegraphics[width=7.0in,angle=0]{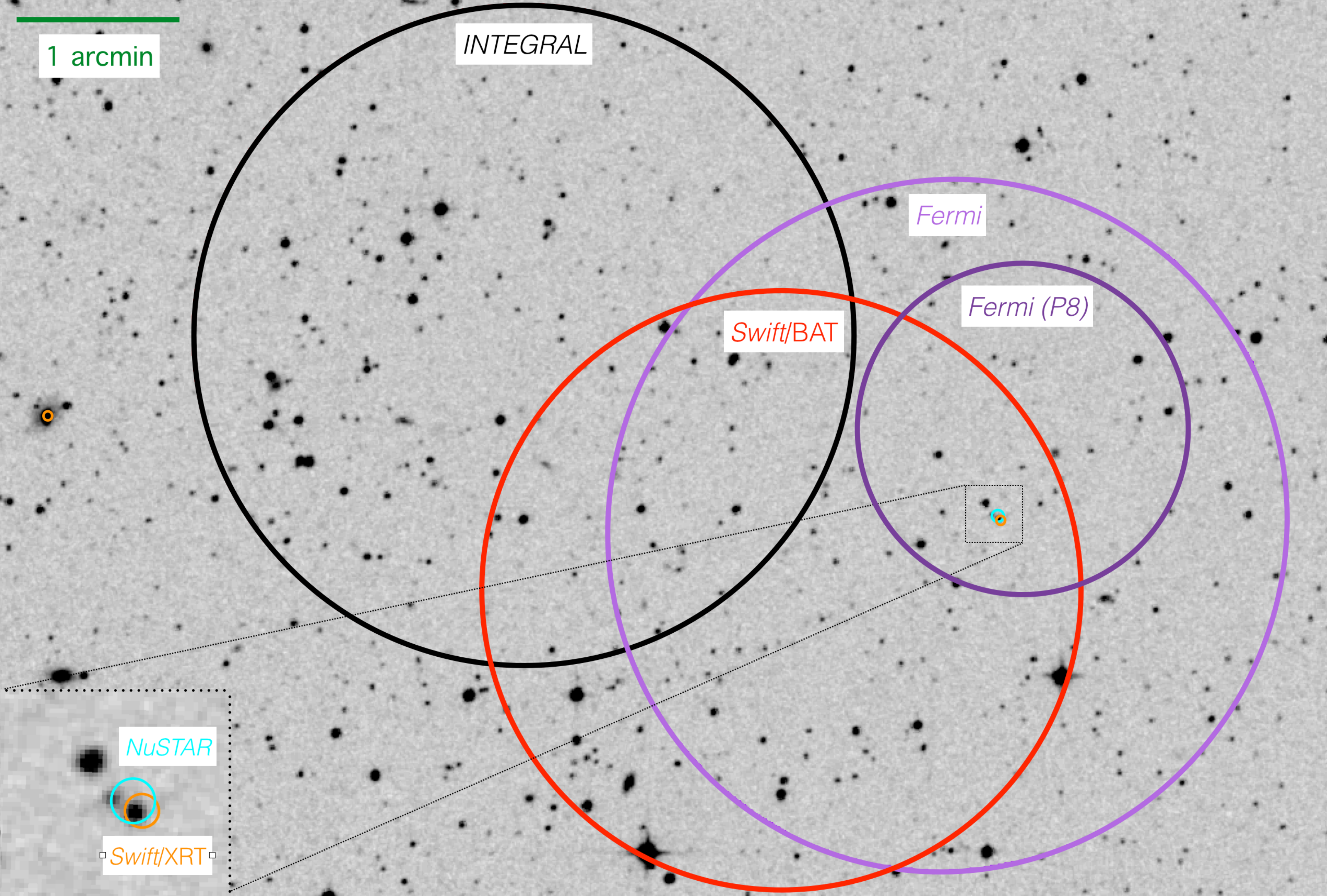}
\caption{Localization of X-ray and $\gamma$-ray sources on a red Digitized Sky Survey image, with typical orientation (N up, E left). The initial \emph{INTEGRAL} detection (black circle) was resolved into two \emph{Swift}/XRT sources (small orange circles). The eastern X-ray source matches a background galaxy and the western an optical point source. \emph{Swift}/BAT (red circle) also detected a blend of these two XRT sources. \emph{NuSTAR} data (cyan) confirm hard X-ray emission from the western \emph{Swift}/XRT source, which sits at the center of the 3FGL \emph{Fermi}-LAT error ellipse (light purple) and within the revised  \emph{Fermi} Pass 8 region (dark purple; \S 3.3.3), strongly suggesting $\gamma$-ray emission from this object. This well-localized optical/X-ray source is the subject of the paper.}
\label{fig:dss}
\end{figure*}

\subsection{Initial Source Discovery and High-Energy Observations}

\subsubsection{{INTEGRAL} and {Swift}} 

 An X-ray source in this region was first detected in the 20--60 keV band by \emph{INTEGRAL} as IGR J04288--6702 (Grebenev et al.~2013). Follow-up  0.3--10 keV \emph{Swift}/XRT follow-up observations resolved this source into two separate sources visible in softer X-rays (Figure \ref{fig:dss}). A hard (14--195 keV) detection in this region was also reported by the \emph{Swift}/BAT 70-month survey (Baumgartner \etal~2013). Here the source is noted as a close ``double" source,  suggesting the possibility that both \emph{Swift}/XRT sources may have been detected by \emph{Swift}/BAT. The \emph{Swift}/BAT 14--195 keV flux is $1.5 \times 10^{-11}$ erg s$^{-1}$ cm$^{-2}$, and the best-fit power law has a photon index $\Gamma = 0.95$. However, this flux value should be interpreted with caution due to the blended nature of the source.
 
The brighter of the \emph{Swift}/XRT sources (1SXPS J042947.1--670320; mean count rate $(5.6\pm0.3) \times 10^{-2}$ cts sec$^{-1}$) has an absorbed power-law with $\Gamma = 1.8\pm0.2$, $N_H$ consistent with the Galactic value of $5 \times 10^{20}$ cm$^{-2}$, and an unabsorbed flux of $2.3\pm0.2 \times 10^{-12}$ erg s$^{-1}$ cm$^{-2}$ (Evans et al.~2014). To study the fainter source (1SXPS J042749.2--670434; mean count rate $(1.2\pm0.2) \times 10^{-2}$ cts sec$^{-1}$) in more detail, we analyzed 10 ksec of  \emph{Swift}/XRT observations taken in 2010 October (P.I.~Markwardt). Using the task \texttt{xrtgrblc} of \texttt{HEAsoft} (v6.19), we extracted the spectrum of the source, which can be described by a hard absorbed power law with parameters of $N_H = 2^{+4}_{-2} \times10^{22}$  cm$^{-2}$, $\Gamma_\mathrm{sw}=1.0_{-1.6}^{+2.1}$, and $F_\mathrm{0.3-10keV}=2.0^{+15.9}_{-0.5}\times10^{-12}$\flux\ ($\chi_\nu^2=2.50/1$; all uncertainties are in 90\% confidence). As only about 80 useful counts were collected, the XRT light curve is too noisy to show any significant variability. We note that an additional \emph{Swift}/XRT observation was taken simultaneously with the \emph{NuSTAR} data (\S 2.1.4), but with only 5 net counts, these data were not useful for analysis, though we do analyze the associated \emph{Swift}/UVOT photometry in \S 2.2.3.

\subsubsection{{Fermi}}

A $\gamma$-ray source in this region is present in the in the \emph{Fermi}-LAT 3FGL catalog (Acero et al.~2015), listed as 3FGL J0427.9--6704, with J2000 right ascension and declination of (04:27:55.7, --67:04:40). The positional error ellipse is nearly circular, with a 95\% radius of $\sim 4.2\arcmin$. The catalog flux of the source is ($9.4\pm0.8$) $\times 10^{-12}$ erg s$^{-1}$ cm$^{-2}$  (0.1--100 GeV). The $\gamma$-ray spectrum is well-fit by a power law with an index of $\Gamma = 2.46\pm0.07$. While the detection $> 10$ GeV is marginal, there is no evidence of curvature, and the flux in this band is consistent with the power law at lower energies. We have also analyzed Pass 8 observations of this region with very similar spectral results and an improved localization (\S 3.3.3).

\subsubsection{Source Analysis and Identification}

As Figure \ref{fig:dss} shows, the 20--60 keV \emph{INTEGRAL} position is close to halfway between the two \emph{Swift}/XRT sources, suggesting they have comparable flux at these energies (and unlike in the soft 0.3--10 keV XRT band, where the count rates are different by a factor of $\sim 5$). The  \emph{Swift}/BAT 14--195 keV position is yet closer to the fainter XRT source. Comparing the location of the \emph{Fermi}-LAT $\gamma$-ray source to the two \emph{Swift}/XRT sources: the fainter source is near the center of the 3FGL $\gamma$-ray error ellipse, while the brighter XRT source sits nearly $\sim 11\arcmin$ away (Figure \ref{fig:dss}), far outside the \emph{Fermi} 3FGL error ellipse.  There are no other significant X-ray sources within the $\gamma$-ray error ellipse. 

All of the high-energy observations are consistent with an interpretation in which there are two X-ray sources in this region with differing spectra, such that the fainter \emph{Swift}/XRT source is also harder. It gradually overtakes the flux of the brighter \emph{Swift}/XRT source at higher energies, and in $\gamma$-rays is the only source detected. In this scenario all of the \emph{Fermi}-LAT flux and much of the blended \emph{Swift}/BAT flux comes from this source, which we call 3FGL J0427.9--6704 henceforth.

The high-resolution \emph{Swift}/XRT data allow the identification of optical counterparts to the X-ray sources. To within 2\arcsec, the \emph{Swift}/XRT position of 3FGL J0427.9--6704  matches a moderately bright point source (J2000 right ascension and declination of 04:27:49.61, --67:04:35.0) with $R\sim 17.1$ (Monet et al.~2003). As we will show in \S 3, this source shows all the optical properties of an accreting Galactic compact binary and is thus nearly certain to be associated with the X-ray and $\gamma$-ray emission discussed here. The brighter, softer \emph{Swift}/XRT source is located at the center of a distant galaxy and is therefore a likely active galactic nucleus.
 
 \subsubsection{NuSTAR}
 
For follow-up observations of the \emph{Swift}/XRT source associated with 3FGL J0427.9--6704, on 2016 May 19, we obtained 84 ksec of observations with \emph{NuSTAR} (Harrison et al.~2013), with 3FGL J0427.9--6704 placed at the nominal aimpoint. Of the 84 ksec, 60 ksec were on-source. 3FGL J0427.9--6704 is clearly detected with a net count rate of $(4.89\pm0.08)\times10^{-2}$ cts s$^{-1}$ per detector (about 6800 source counts in total). These data are analyzed in \S 3.3.

 \subsection{UV, Optical and Near-IR Photometry}

\subsubsection{OGLE}

While the location of the source is west of the main body of the LMC, it is covered in the ongoing OGLE-IV survey, which has added fields to monitor the bridge area between the Magellanic Clouds (Udalski et al.~2015).  We converted the OGLE instrumental magnitudes $i$ and $v$ into standard $I$ and $V$ filters using the equations $I = i - 0.338 - 0.0047 (v-i)$ and $V = v - 0.066 - 0.072 (v-i)$, with $(v-i)$ assumed from a smoothed version of the light curve for epochs with only one filter available. We also converted the OGLE UTC HJD values into Barycentric Julian Date (BJD) on the Barycentric Dynamical Time (TDB) system (Eastman et al.~2010). The OGLE photometry comprise 368 epochs in $I$ and 28 epochs in $V$, spanning 07 March 2010 to 29 November 2015.

\subsubsection{SMARTS}

From 16 Jan to 1 May 2016  (UT) we obtained optical $BVI$ and near-IR $H$ photometry of the source on most clear nights using ANDICAM on the SMARTS 1.3-m telescope at CTIO. On a few nights multiple sets of observations were taken; the total number of epochs was 62--64 per band, depending on the band. The exposure time in each of $BVI$ was 300 sec; the on-source exposure time in $H$ varied but was typically 500--600 sec. The data were reduced as described in Walter et al.~(2012). We performed differential aperture photometry of the source with respect to ten ($BVI$) or four ($H$) bright stars in the field, with absolute calibration via Landolt (1992) standards on clear nights ($BVI$) or with respect to 2MASS ($H$). All the OGLE and SMARTS photometry can be found in Table \ref{tab:bigg}.

\begin{deluxetable}{crrr}
\tablecaption{OGLE and SMARTS Photometry of 3FGL J0427.9--6704 \label{tab:bigg}}
\tablehead{BJD & Vega Mag & Err. & Band \\
                   (d)  & (mag) & (mag) & }

\startdata
2457404.5505499 & 18.068 & 0.018 & B \\
2457404.6303346 & 18.468 & 0.026 & B \\
2457408.5553331 & 18.552 & 0.032 & B \\
2457408.5988911 & 18.333 & 0.025 & B \\
2457408.6483436 & 18.624 & 0.031 & B \\
2457409.5810375 & 18.406 & 0.037 & B \\
... \\
\enddata
\tablecomments{Table 1 is published in its entirety in the machine-readable format. A portion is shown here for guidance regarding its form and content.
These magnitudes are on the Vega system and not corrected for extinction.}
\end{deluxetable}

\begin{deluxetable}{lr}
\tablecaption{Mean Panchromatic Photometry\tablenotemark{a} \label{tab:pan}}
\tablehead{Filter    & AB mag }
\startdata
$uvw2$ & $19.84\pm0.17$ \\ 
$uvm2$ & $19.77\pm0.03$ \\ 
$uvw1$ & $19.44\pm0.12$ \\  
$u$    & $18.51\pm0.22$ \\     
$B$    & $18.01\pm0.03$ \\
$V$    & $17.72\pm0.03$ \\
$I$    & $17.27\pm0.01$ \\  
$J$    & $17.05\pm0.05$ \\   
$H$    & $17.13\pm0.02$ \\ 
$K$    & $17.28\pm0.09$ \\   
$W1$   & $18.25\pm0.03$ \\ 
$W2$   & $18.75\pm0.06$ \\ 
\enddata
\tablenotetext{a}{For bands with multiple observations, the mean magnitudes out of eclipse (taken as $\phi \pm 0.05$ of each eclipse) and standard errors of the mean are given. These
magnitudes are not corrected for extinction.}
\end{deluxetable}

\subsubsection{UV and Near-IR}

We obtained UV photometry from \emph{Swift}/UVOT observations of the field. In the 2010 and 2016 \emph{Swift} observations, the object is detected in all of $uvw1$, $uvm1$, $uvw2$, and $u$.  For the near-IR, we retrieved 2MASS $JK$ and WISE $W1$/$W2$ measurements from the 2MASS 6x  (Cutri et al.~2012) and AllWISE (Cutri et al.~2014) catalogs, respectively. To facilitate modeling of the mean spectral energy distribution of the system, we list the mean out-of-eclipse magnitudes for all bands in Table \ref{tab:pan}. We emphasize that any such modeling should be undertaken with the knowledge that the UV, optical, and IR emission is variable and that our UV and IR data (excepting $H$)  only represent one to a few epochs.

\subsection{Optical Spectroscopy}

We performed optical spectroscopy of the source using the Goodman Spectrograph (Clemens et al.~2004) on the SOAR telescope from 2015 August 23 to 2016 March 26. Observations were made either with a 1.03\arcsec\ slit and 400 l mm$^{-1}$ or 1200 l mm$^{-1}$ grating, with exposure times ranging from 5 to 20 min. These observations cover both short and long timescales, with 22 individual low-resolution spectra and 97 medium-resolution spectra. The low-resolution spectra cover an usable wavelength range of $\sim 3400$--7000 \AA\ at a resolution of 5.6 \AA. The 1200 l mm$^{-1}$ observations span 4300--5600 \AA\ at a resolution of 1.7 \AA, with the wavelength range chosen to contain both strong emission and absorption lines. The spectra were reduced and optimally extracted in the usual manner.

\section{Results and Discussion}

\subsection{An Eclipsing Binary}

The $I$-band data, mostly from OGLE, span more than 6 years and offer the best long-term picture of the optical source. A time series of these data show substantial scatter---much more than expected on the basis of photometric uncertainties---along with regular dimming by more than 1 mag. This immediately suggested the system might be an eclipsing binary. To explore this further, we initially searched for a photometric period using a Lomb-Scargle periodigram, finding a strong peak around 0.3667 d. We refined this period determination through a $\chi^2$ minimization of a SuperSmoother fit to the data (Friedman 1984) for varying periods. The period derived in this manner is $P = 0.3667200(7)$ d, with the uncertainty calculated via bootstrap. The BJD at mid-eclipse is 2455912.83987(95).

\begin{figure}[t]
\includegraphics[width=3.4in]{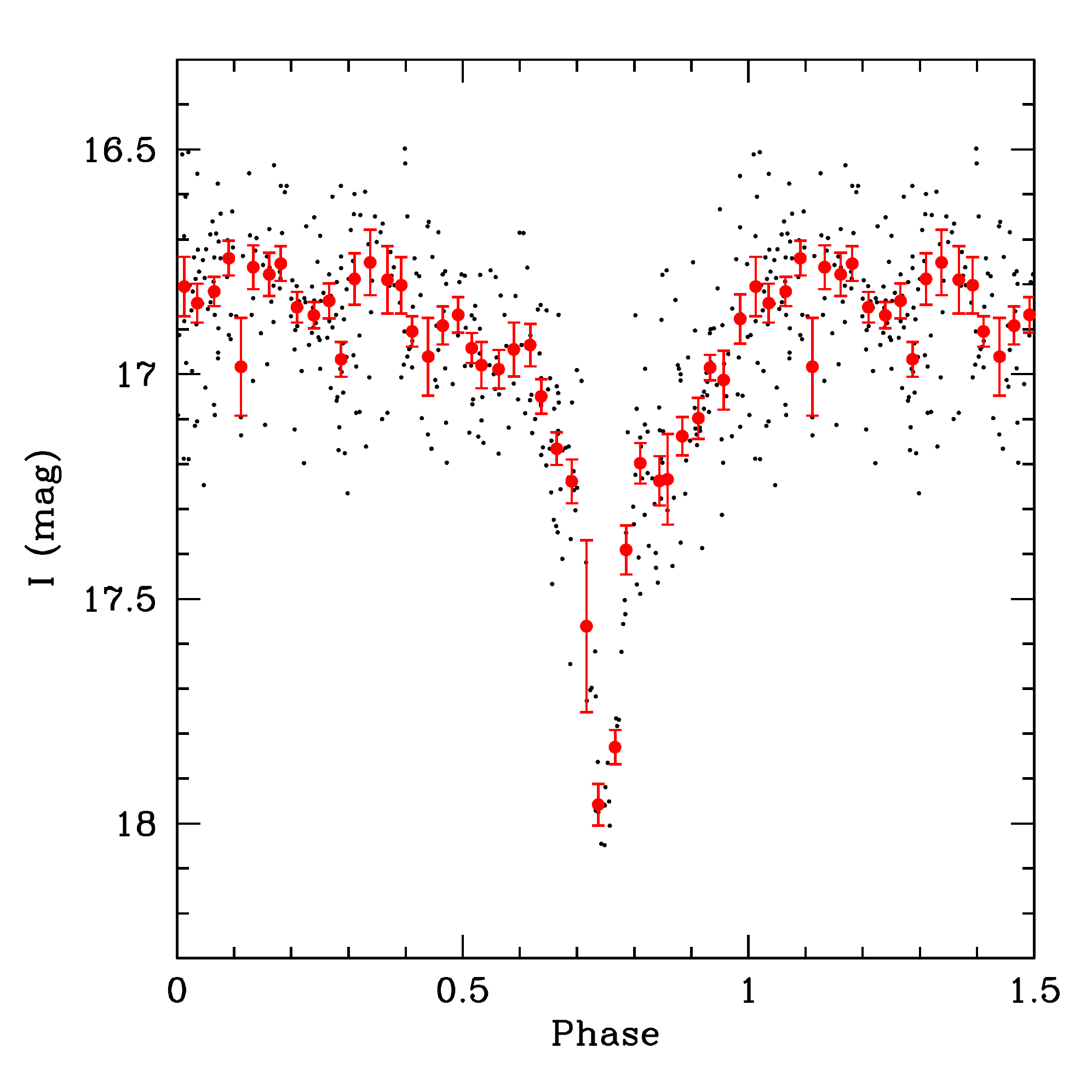}
\caption{$I$-band light curve phased on the orbital period of $P = 0.3667200$ d. The small black circles are the original photometry and the red circles with uncertainties a binned light curve. Both the deep primary eclipse ($\phi = 0.75$) and the secondary eclipse  ($\phi = 0.25$) are clearly visible.}
\label{fig:lc_opt}
\end{figure}

The phased light curve is shown in Figure \ref{fig:lc_opt}. The source is clearly an eclipsing binary, with a deep eclipse of $> 1$ mag in $I$. Outside of eclipse, the light curve shows a large scatter. The individual uncertainties on the photometric data points are typically 0.01 mag, while the rms dispersion out of the main eclipse is about 0.16 mag. This scatter is therefore not due to random uncertainties but to real variations in the total flux of the system. In the context of our interpretation of the system as a low-mass X-ray binary (\S 3.3), these variations are likely due to flickering from the accretion disk, possibly related to reprocessing of a rapidly varying X-ray flux. While we do not have short-term time-series photometry of the system, evidence from spectroscopy (\S 3.2) suggests that large-amplitude flickering is occurring on time scales as short as 5 min, consistent with the timescale on which the X-ray flux is observed to vary.

Hidden within the photometric scatter and separated in phase by 0.5 from the main eclipse, there is a much weaker secondary eclipse that is more apparent in the binned light curve (Figure \ref{fig:lc_opt}). Overall, the system is also brighter around the secondary eclipse than the primary eclipse, indicative of the reflection effect due to reprocessed radiation from the X-ray source. In \S 3.4 we perform more detailed physical modeling of the light curves and show phased light curves in other photometric bands.

\subsection{Analysis of Spectroscopy}

\begin{figure}[t]
\includegraphics[width=3.4in]{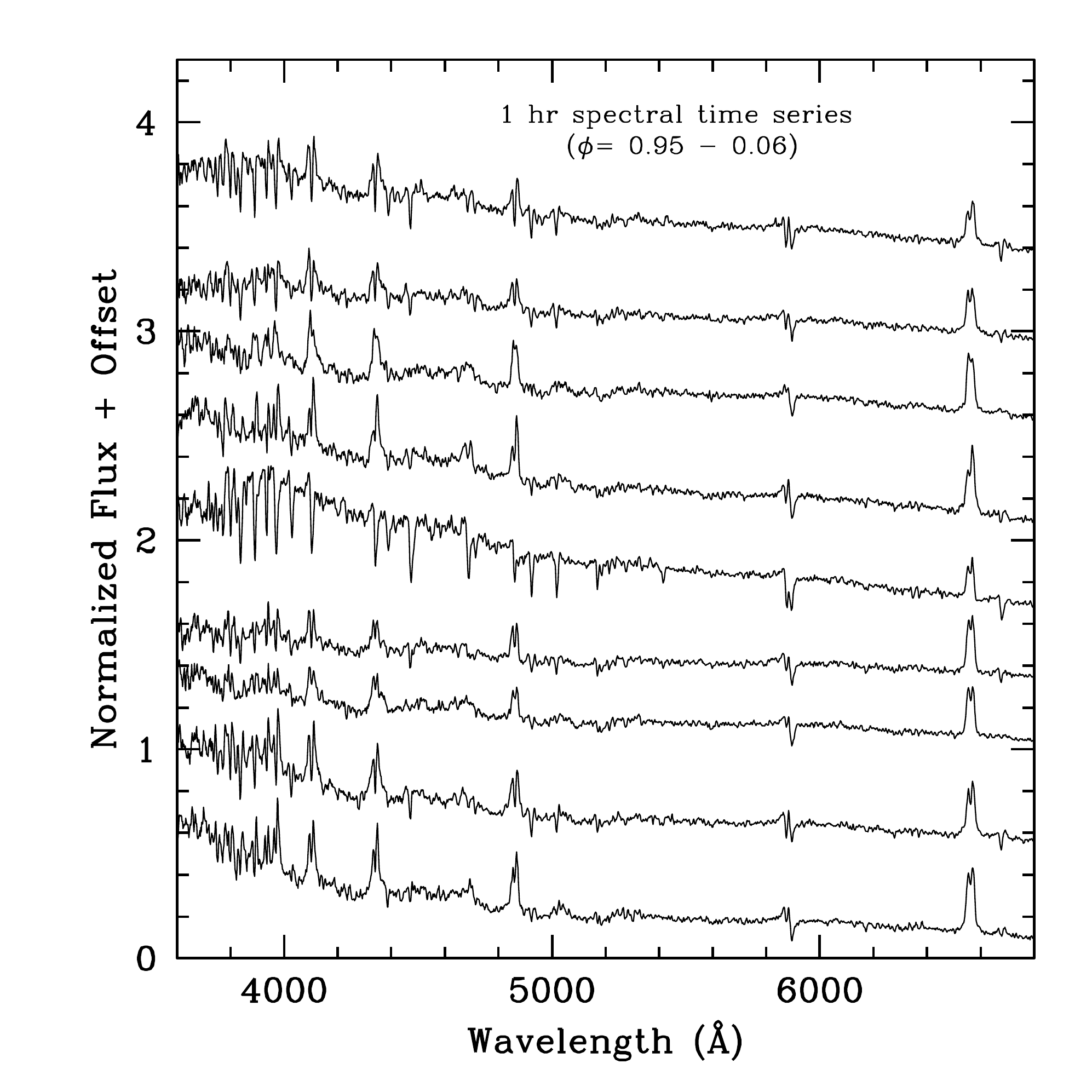}
\caption{Time series of low-resolution optical spectra of the system from 30 January 2016 covering 1 hr, with phases between $\phi = 0.95$ and 0.06 (bottom to top, respectively), chosen to be far from eclipse. The spectra show enormous variations in the prominence of the emission-line contribution. This is most notable in the middle spectra, in which the spectrum switches from one with moderate emission lines to one strongly dominated by absorption lines in 5 min, and then to just as strong domination by emission lines just 5 min later.}
\label{fig:spec}
\end{figure}

The identification of an eclipsing binary with associated high-energy emission points to a classification as an X-ray binary powered by accretion onto a compact object. Our SOAR spectroscopy of the source confirms this scenario. The low-resolution spectra are remarkably variable, with the changes apparently dominated by the disk flickering (Figure \ref{fig:spec}).

Here we describe the basic features apparent after examination of the low and medium-resolution spectra. At most epochs, there is strong double-peaked Balmer emission. When the emission is strongest the Balmer emission can be seen down to at least H8 (see several spectra in Figure \ref{fig:spec}). The velocity separation of the peaks depends on the epoch and which Balmer line is being considered. Taking H$\beta$, for which we have the best data, the distribution of peak separations has a median of $1216\pm23$ km s$^{-1}$ and $\sigma = 175$ km s$^{-1}$. This distribution is much broader than expected solely from the uncertainties on the peak locations (see below) and so likely represents true variations. Each set of emission peaks has central Balmer absorption, which in some cases is deeper than the local continuum. High-order Balmer lines are visible in absorption to at least $n=14$ in spectra where the contribution from emission lines is smaller (Figure \ref{fig:spec}; middle spectrum).
 
Setting aside the Balmer lines, there are other observable absorption lines. Ignoring the (possibly interstellar) sodium resonance doublet, the next strongest absorption line is \ion{He}{1} at 4471 \AA. The other strong \ion{He}{1} lines (e.g., 6678 \AA) are clearly detected. In the emission-dominated epochs, \ion{He}{1} is also seen clearly in emission around 
4026 \AA, 5876 \AA, and 6678 \AA. In all cases, this \ion{He}{1}  emission is much weaker than the Balmer emission. We also see \ion{He}{2} absorption at 4686 \AA\ and 5412 \AA, though none at 4542 \AA. In a few spectra there is clear evidence for \ion{He}{2} at 4686 \AA\ in emission.

If we associate the absorption lines with the companion star to the compact object, we could use them to constrain the spectral type of the donor, keeping in mind that the strong, broad emission lines obviate the possibility of a simple quantitative analysis. The absorption spectrum can generally be described as having strong hydrogen and \ion{He}{1} absorption with some \ion{He}{2} absorption, which would place it at the border of the O and B stellar classes with a rough effective temperature of around 32000 K. However, we argue in the next subsection that these strong optical absorption lines are associated with the \emph{disk} rather than the secondary. 

In Figure \ref{fig:spec} we plot a series of low-resolution spectra taken over 1 hr on 30 January 2016, covering phases $\phi = 0.95$--0.06, well away from the eclipses. These spectra show the extremes of the absorption line and emission line contributions. The variability of the spectra is also notable, with a switch between these extremes on a timescale as short at 5 min.

\subsubsection{Radial Velocities}

\begin{figure*}[t]
\includegraphics[width=7.0in]{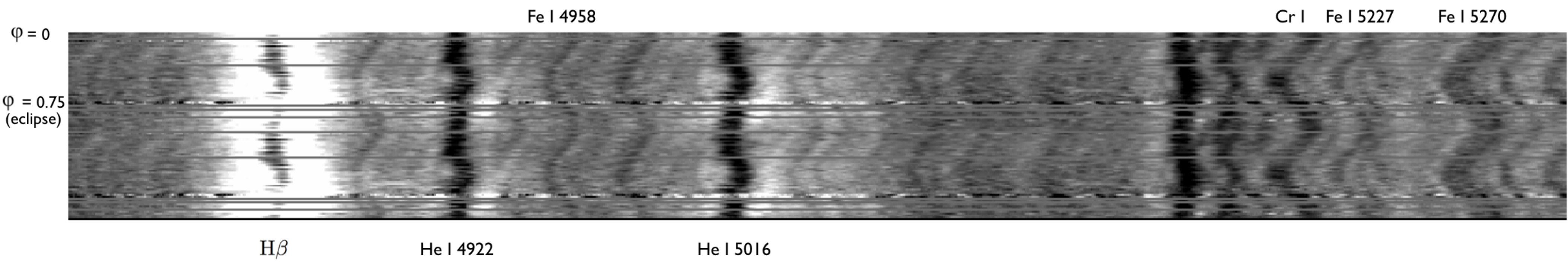}
\caption{Medium-resolution optical spectra of the system binned in phase, going from $\phi = 0$ to 2 top to bottom (i.e., with all phases repeated). The wavelength range is 4790--5300 \AA. The strongest absorption lines match the variations in the double-peaked H$\beta$ emission in velocity amplitude and phase. However, many weaker lines, offset in phase by $0.5$ and of higher velocity amplitude, are also visible. We associate the emission and strong absorption lines with the accretion disk and the weak absorption lines with the secondary. Some of the individual lines in the regions used to determine the radial velocities are marked.}
\label{fig:trailed}
\end{figure*}

Given that the eclipsing nature of the system implies a close to edge-on orientation, velocity measurements could in principle allow a good constraint on the mass of the compact object primary in the binary. Therefore we obtained a large number of spectra with the medium-resolution 1200 l mm$^{-1}$ grating to measure precise radial velocities for the secondary. As we explain in detail below, the results were unexpected.

Figure \ref{fig:trailed} shows a trailed spectrograph of medium-resolution spectra over the wavelength range 4790--5300 \AA\ in bins of 0.02 in phase and smoothed by a 5-pixel boxcar. The strong double-peaked H$\beta$ emission and deep \ion{He}{1} absorption lines at 4922 and 5016 \AA\ are the most obvious features (absorption associated with Mg$b$ around 5170 \AA\ also appears to be present). Qualitatively, these absorption lines match the H$\beta$ emission in both velocity amplitude and phase. However, there is a forest of weaker absorption lines that were not obvious in the examination of the low-resolution spectra. These weaker lines have a much larger radial velocity amplitude and are offset in phase from the strong absorption and H$\beta$ emission lines. These findings lead to the unexpected conclusion that the strong absorption lines are associated with the \emph{disk} rather than the secondary, whose motion is instead traced by the weaker absorption lines.

To put these conclusions on a more quantitive basis, we measured the strong absorption-line radial velocities using the \ion{He}{1} lines at 4922 \AA\ and 5016 \AA, which are the strong lines least affected by emission or confusion with other lines. The velocities were determined through cross-correlation with a spectrum of the B3V star HD 210121 taken with the same instrumental setup. To measure the weaker lines, we selected the cleanest, most isolated regions containing these lines, which we judged to be 4950--4970 \AA\ and 5195--5285 \AA. Of these lines, several can be clearly attributed to \ion{Fe}{1} and \ion{Cr}{1}, suggesting a mid-to-late K spectral type of the secondary. We determined the radial velocity from these weak lines through cross-correlation with a spectrum of the K7V star HD118100.

Our spectral range was chosen to contain H$\beta$ and nearly all spectra show double-peaked H$\beta$ emission likely formed in the outer regions of the accretion disk. To measure the orbital motion of this disk, we performed a three-component fit to the spectrum in the region of H$\beta$, fitting the background and two Gaussian line components. In general the approaching (blue) component of the line is somewhat weaker and slightly asymmetric and hence we focus on the receding (red) line for comparison with the absorption-line radial velocities. 

\begin{figure}[t]
\includegraphics[width=3.4in]{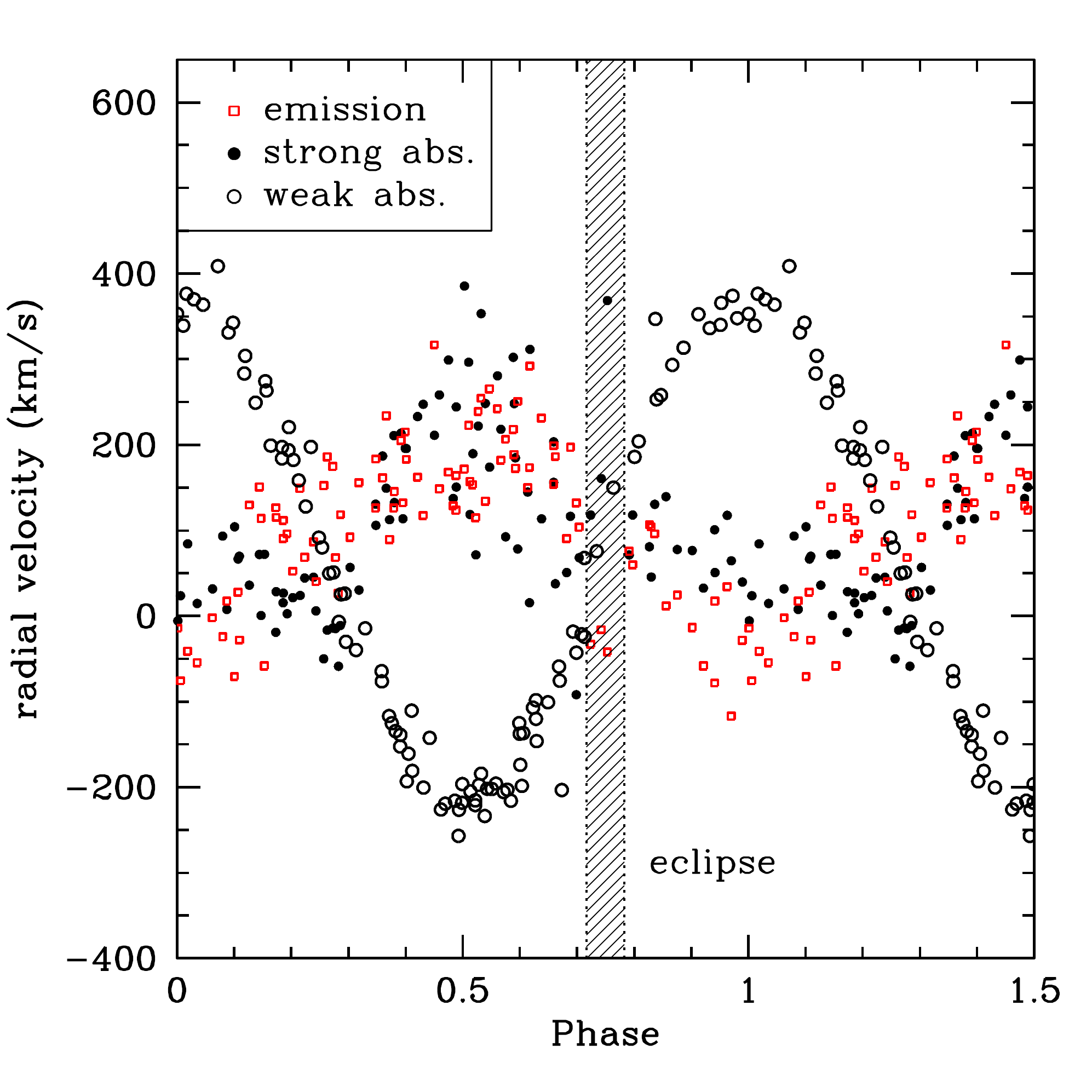}
\caption{Phased radial velocity curves of the strong absorption lines (black filled circles), weak absorption lines (black open circles), and the receding component of the H$\beta$ emission (red unfilled squares), with the latter offset by --500 km s$^{-1}$. The X-ray eclipse region is shaded. The phase convention is such that $\phi = 0.75$ is the superior conjunction of the compact object, matching the X-ray eclipse. Both the emission lines and strong absorption lines peak at $\phi=0.5$, implying these absorption lines are associated with the disk rather than the secondary. The radial velocities of the weak lines trace a clean circular Keplerian curve offset in phase by 0.5, indicating that they trace the secondary star.}
\label{fig:rv}
\end{figure}

Figure \ref{fig:rv} shows the strong and weak absorption-line and red emission-line radial velocities plotted as a function of orbital phase, where we have anticipated the results of \S 3.3 and assumed that the $I$-band eclipse corresponds to superior conjunction of the compact object, which we set as $\phi = 0.75$ (in this phase convention the ascending node of the compact object is $\phi = 0.5$). To better compare the emission-line to strong absorption line velocities, we have added an offset to the emission line velocities as only the red component of the H$\beta$ emission is plotted. Outside of eclipse, this red component is consistent with a circular Keplerian orbit peaking at $\phi \sim 0.5$, as expected if the disk surrounds the compact object that is eclipsed at $\phi = 0.75$. 

Remarkably, the strong absorption line radial velocities \emph{also} peak at $\phi \sim 0.5$, matching the emission. This is wholly inconsistent with the natural association of absorption lines with the secondary, whose lines would be expected to peak at $\phi = 0$. The interpretation most consistent with these data is that the strong absorption lines are associated with the disk, not the secondary, matching our initial finding on the basis of the trailed spectrogram in Figure \ref{fig:trailed}. We note that the absorption line radial velocity curve does not have a simple Keplerian shape, with a minimum closer to $\phi \sim 0.25$ than 0 as expected in our proposed scenario, and a quick rise to the maximum at $\phi \sim 0.5$. 

The strong absorption lines appear resolved, with a typical full-width at half-maximum of $\sim 300$--400 km s$^{-1}$, much narrower than the emission lines (\S 3.2 and \S3.4). This is consistent with the formation of these absorption lines in a restricted region of the cool outer disk along the line of the sight to the hot inner disk. However, the detailed formation of these unusual absorption lines (apparent at nearly all epochs) may be complex and we do not model them further in this paper. To our knowledge, optical absorption lines have never previously been observed  in a low-mass X-ray binary accreting in the low state.

In contrast to the rather noisy radial velocity curves of the emission and strong absorption lines, the velocities of the weak lines show a clean circular Keplerian curve that peaks at $\phi = 0$, exactly out of phase with the other curves. This confirms that these lines arise in the faint secondary and trace its motion around the binary's center of mass, and indeed that this system is the equivalent of an eclipsing double-lined spectroscopic binary.

To characterize these orbits, we fit circular Keplerian models to the radial velocities with the period and time of conjunction fixed to the values determined from the photometry. Focusing first on the red emission line radial velocities: as  the rms at each phase is evidently much larger than expected on the basis of the formal uncertainties, we first did an unweighted fit, finding a semi-amplitude $K_{em} = 103\pm8$ km s$^{-1}$ and a systematic velocity (for this line) of $v_{\rm red} = 596\pm6$ km s$^{-1}$. This fit had an rms of 52  km s$^{-1}$ (compared to a mean formal uncertainty of 21 km s$^{-1}$). A weighted fit yields $K_{em} = 106\pm5$ km s$^{-1}$ and $v_{\rm red} = 592\pm4$ km s$^{-1}$, with an rms of 54 km s$^{-1}$. As the relative uncertainties probably do capture information about the fitting, it may be that the best estimate lies between these listed values, but for the remainder of the paper we assume the former, unweighted values.

As noted above, the absorption line radial velocity curve is not well-fit by a Keplerian curve in the range $\phi \sim 0.2$--0.4. Nonetheless, we performed such a fit as a check on the emission-line model. The fit rms is higher as expected (70 km s$^{-1}$), however, the best-fit $K_{em} = 99\pm11$ km s$^{-1}$, entirely consistent with that from the emission-line fit.

For the weak absorption lines that trace the secondary, we find the secondary semi-amplitude $K_2 = 293\pm4$ km s$^{-1}$ and $v_{sys} = 79\pm3$ km s$^{-1}$. The rms of this fit is 14 km s$^{-1}$, much smaller than that for disk velocities, but still somewhat higher than expected on the basis of the individual uncertainties, probably because of a varying disk contribution to the regions used for cross-correlation. One potential concern is that $K_2$ is biased low from the observed center-of-mass motion due to irradiation of the secondary. This can appear as a false eccentricity in the velocity curve (e.g., Davey \& Smith 1992). To test whether this is occurring, we tried allowing a non-circular fit, but found that this did not improve the fit. As a second test, we fit the radial velocity curves of the warm ($\phi = 0$--0.5) and cold ($\phi = 0.5$--1) sides of the secondary separately, finding $K_2$ values consistent to within 1\%. In addition, the strength of the weak lines does not appear to appreciably vary over the orbit. We conclude that there is no evidence that $K_2$ is significantly affected by irradiation.

\subsection{The High-Energy Emission}

\subsubsection{\emph{NuSTAR} X-ray Spectrum}

\begin{figure*}[t]
\centering
\includegraphics[width=7in]{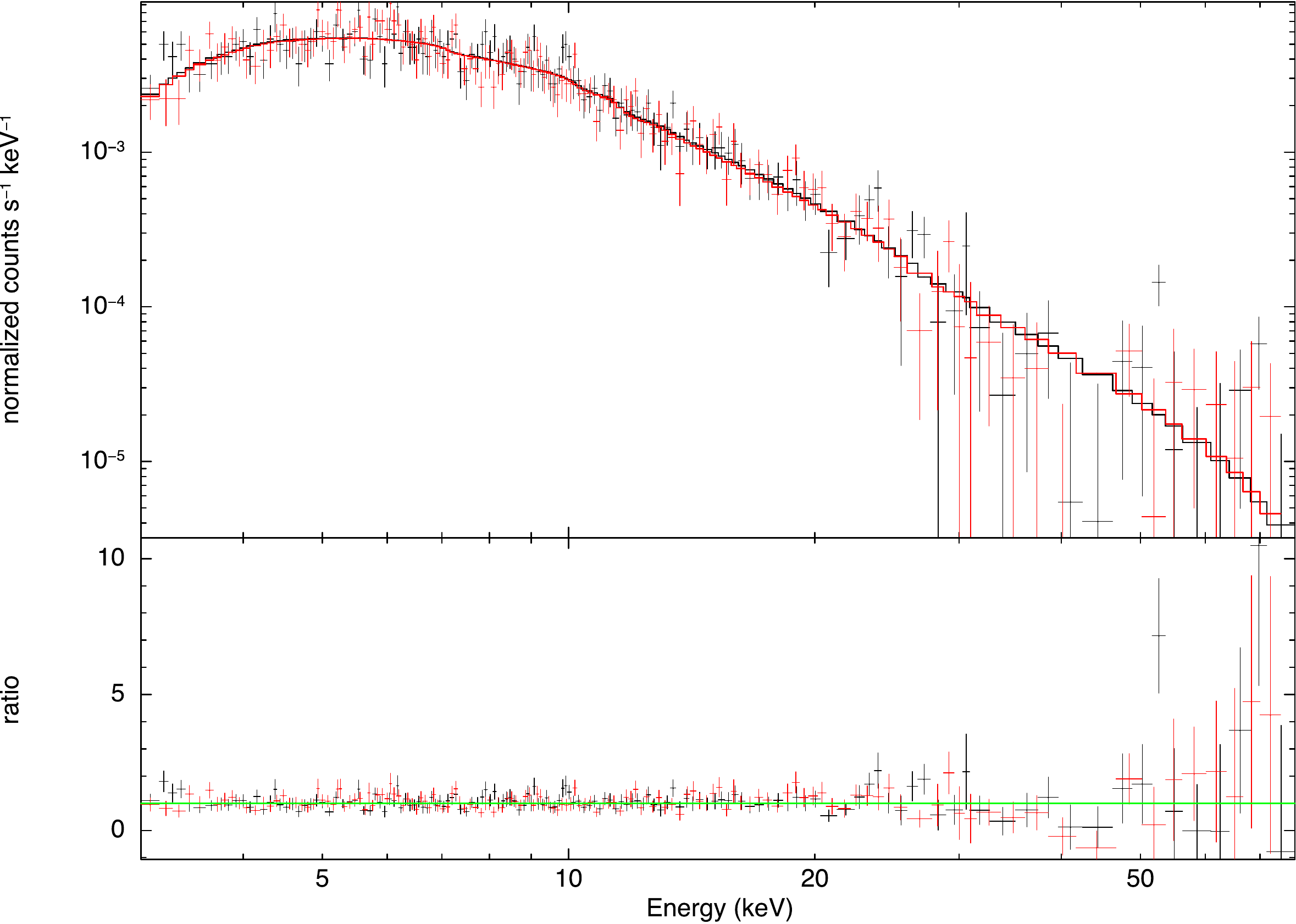}
\caption{The \textit{NuSTAR} FPMA (black) and FPMB (red) spectra over 3--79~keV fit with an absorbed power-law.}
\label{fig:nu_spec}
\end{figure*}

We analyzed the 84 ksec of \emph{NuSTAR} data using the \texttt{nuproducts} task of \texttt{HEAsoft} with a 60\arcsec\ radius circular extraction region and a source-free 60\arcsec\ background region, extracting barycentric-corrected light curves and spectra over the energy range 3--79 keV using both focal plane modules A and B (FPMA/B). After spectral binning the {\emph{NuSTAR} data with a minimum of 20 counts per bin, we fit them with an absorbed power-law (i.e., \texttt{const*phabs*cflux*pow} in \texttt{XSPEC}, using a multiplicative factor \texttt{const} for the cross-calibration between FPMA and FPMB: $\mathrm{C_{FPMB}}=F_\mathrm{FPMB}/F_\mathrm{FPMA}$). All fits used the abundance scale of Anders \& 
 Grevesse (1989). The best-fit parameters are $\mathrm{C_{FPMB}}=1.05\pm0.05$, \nh\ $=5.8^{+1.8}_{-1.7}\times10^{22}$\cm, $\Gamma =1.68^{+0.09}_{-0.08}$, and $F_\mathrm{3-79keV}=(6.7\pm0.4)\times10^{-12}$\flux\ ($\chi_\nu^2=254.0/280$). Both the photon index and the column density are consistent with that of the \emph{Swift}/XRT spectrum taken in October 2010, though with much smaller uncertainties. We note that this best-fit flux includes the X-ray eclipse (\S 3.3.2) and so the best estimate of the uneclipsed flux is 6\% higher: $F_\mathrm{3-79keV}=(7.1\pm0.4)\times10^{-12}$\flux. Given the low foreground extinction, the high \nh\ is likely due to material associated with the disk that produces photoelectric absorption due to the high inclination of the system.

As a check on the consistency between the \emph{NuSTAR} and \emph{Swift} data, we also fit the 2010 \emph{Swift} and 2016  \emph{NuSTAR} X-ray spectra together with an additional cross-calibration factor $\mathrm{C_{XRT}}=F_\mathrm{XRT}/F_\mathrm{FPMA}$. The results are $\mathrm{C_{FPMB}}=1.06\pm0.05$, $\mathrm{C_{XRT}}=0.9\pm0.2$, \nh\ $=4.4^{+1.6}_{-1.2}\times10^{22}$\cm, $\Gamma =1.63^{+0.08}_{-0.07}$, and $F_\mathrm{3-79keV}=(6.8\pm0.4)\times10^{-12}$\flux\ ($\chi_\nu^2=260.3/283$). All of the best-fit parameters still remain within the original \emph{NuSTAR}-only uncertainties. The \emph{Swift}--\emph{NuSTAR} cross-calibration factor is $\sim 0.9$, consistent with no substantial X-ray flux change between the epochs. Nonetheless, given the possibility for modest variations in the system and the relatively low net counts in the \emph{Swift} data, we only fit the \emph{NuSTAR} data in the following  spectral analysis.

Given the high column density (\nh\ $>10^{22}$\cm) it is difficult to constrain the presence of an additional minor thermal component in the spectrum. Formally, the addition of a blackbody component slightly improves the fit: a component with  $T_\mathrm{bb}=5.6^{+1.8}_{-1.7}$~keV and $F_\mathrm{3-79keV}=1.0^{+0.9}_{-0.8}\times10^{-12}$\flux\ results in $\Delta\chi^2=-4.3$ with 2 extra degrees of freedom. However, on the basis of $10^4$ Monte Carlo simulations of the best-fit power-law model, this $\Delta\chi^2$ improvement occurs by chance in about 12\% of simulations, so the evidence for the blackbody component is not significant. Furthermore, this component is not straightforward to interpret; it is much harder, for example, than would be expected for thermal emission from the neutron star surface (e.g., Rutledge et al.~1999).

The X-ray spectrum also shows marginal evidence for an absorption feature at about 40~keV (Figure \ref{fig:nu_spec}). An additional Gaussian absorption line (represented by  \texttt{gabs} in \texttt{XSPEC}) centered at $E_\mathrm{abs}=40^{+4}_{-3}$~keV improves the fit by $\Delta\chi^2=-11.3$ (with 3 additional degrees of freedom). The line width and optical depth are poorly constrained, with 90\% upper limits of $\sigma_\mathrm{abs}<8$~keV and $\tau_\mathrm{abs}>8$, respectively. Using a similar set of spectral simulations, the significance of this improvement is just under 2\%, equivalent to a significance of $\sim2\sigma$. This absorption line does not have a straightforward interpretation, and at its low significance, we believe it is unlikely to be real.

\setlength{\tabcolsep}{1.5pt}
\begin{table*}[ht]
\scriptsize
\centering 
\caption{X-ray spectral fitting results}
\begin{tabular}{@{}lcccccccc}
\hline
Instr.\footnotemark[1] \& Model & $\mathrm{C_{FPMB}}$ & $\mathrm{C_{XRT}}$ & \nh\ & $\Gamma$ & $F_{X,\mathrm{pow}}$\footnotemark[2]  & $T_\mathrm{bb}$ & $F_{X,\mathrm{bb}}$\footnotemark[2]  &  $\chi^2/dof$\\
&&& ($10^{22}$\cm) &  & ($10^{-12}$\flux) & (keV) & ($10^{-12}$\flux) & \\
 \hline
 \hline
 \\
SW (PL) &\nodata&\nodata& $2^{+4}_{-2}$ & $1.0^{+2.1}_{-1.6}$ &$2.0^{+15.9}_{-0.5}$&\nodata&\nodata&2.5/1\\
NU (PL) &$1.05\pm0.05$&\nodata& $5.8^{+1.8}_{-1.7}$ & $1.68^{+0.09}_{-0.08}$ &$6.7\pm0.4$&\nodata&\nodata& 254.03/280\\
NU+SW (PL) &$1.06\pm0.05$&$0.9\pm0.2$& $4.4^{+1.6}_{-1.2}$ & $1.63^{+0.08}_{-0.07}$ &$6.8\pm0.4$&\nodata&\nodata&260.31/283\\
NU (PL+BB)\footnotemark[3] &$1.06\pm0.05$&\nodata& $6.5^{+2.8}_{-2.4}$ & $1.92^{+0.37}_{-0.24}$ &$4.7^{+1.6}_{-1.3}$&$5.6^{+1.8}_{-1.7}$&$1.0^{+0.9}_{-0.8}$& 249.77/278\\
NU (GABS*PL)\footnotemark[4] &$1.06\pm0.05$&\nodata& $4.9^{+1.9}_{-1.8}$ & $1.61\pm0.09$ &$7.0\pm0.5$&\nodata&\nodata& 242.70/277\\
NU high-flux (PL) &$1.05^{+0.07}_{-0.06}$&\nodata& $4.8^{+2.3}_{-2.1}$ & $1.72\pm0.11$ &$11.8\pm0.09$&\nodata&\nodata& 139.84/158\\
NU low-flux (PL) &$1.06\pm0.08$&\nodata& $7.7^{+3.3}_{-3.0}$ & $1.63^{+0.13}_{-0.14}$ &$4.5^{+0.5}_{-0.4}$&\nodata&\nodata& 113.86/142\\
NU high-flux (PL) &$1.05^{+0.07}_{-0.06}$&\nodata& 5.8 (fixed) & $1.76\pm0.06$ &$11.7^{+0.09}_{-0.08}$&\nodata&\nodata& 140.33/159\\
NU low-flux (PL) &$1.06\pm0.08$&\nodata& 5.8 (fixed) & $1.56\pm0.08$ &$4.6^{+0.5}_{-0.4}$&\nodata&\nodata&114.88/143\\
\hline
\end{tabular}
\footnotetext{SW=\emph{Swift}; NU=\emph{NuSTAR}}
\footnotetext{The energy range is 0.3--10~keV for \textit{Swift}/XRT and 3--79~keV for \textit{NuSTAR}. All fluxes are as measured averaged over the orbit, and should be increased by 6\% for the mean out-of-eclipse value.}
\footnotetext{The significance of the improvement over a simple power-law model is about 1$\sigma$. }
\footnotetext{This model also includes an absorption line with $E_\mathrm{abs}=40^{+4}_{-3}$~keV,  $\sigma_\mathrm{abs}<8$~keV, $\tau_\mathrm{abs}>8$. The significance of the improvement over a simple power-law model is about 2$\sigma$. \\}

\label{tab:xray_spec}
\end{table*}

Given the substantial short-term variability (\S 3.3.2), we performed an intensity-resolved spectral analysis to determine whether the photon index varies with X-ray flux. \texttt{XSELECT} was used to resolve the photon events into two roughly-equal populations based on a 100s-bin source light curve: a ``high-flux" group (total FPMA/FPMB source count $>0.13$ cts s$^{-1}$) and a ``low-flux" group below this limit. The spectra of both groups can still be well-described by an absorbed power-law. With $N_H$ free, the high-flux spectrum has a lower \nh\ and a softer photon index than the low-flux spectrum. We also fit fixed $N_H$ models using the best-fit value of the average spectrum (\nh $=5.8\times10^{22}$\cm). We found a photon index of $\Gamma_\mathrm{H}=1.76\pm0.06$ in the high state and $\Gamma_\mathrm{L}=1.56\pm0.08$ in the low state, giving a formal difference of $2\sigma$ between the photon indices. Thus these data are consistent with either a variation in $N_H$ over the orbit or with a changing photon index. 

All the spectral fitting results are summarized in Table \ref{tab:xray_spec}. \\

\subsubsection{X-ray Light Curve}

\begin{figure*}[t]
\centering
\includegraphics[width=7in]{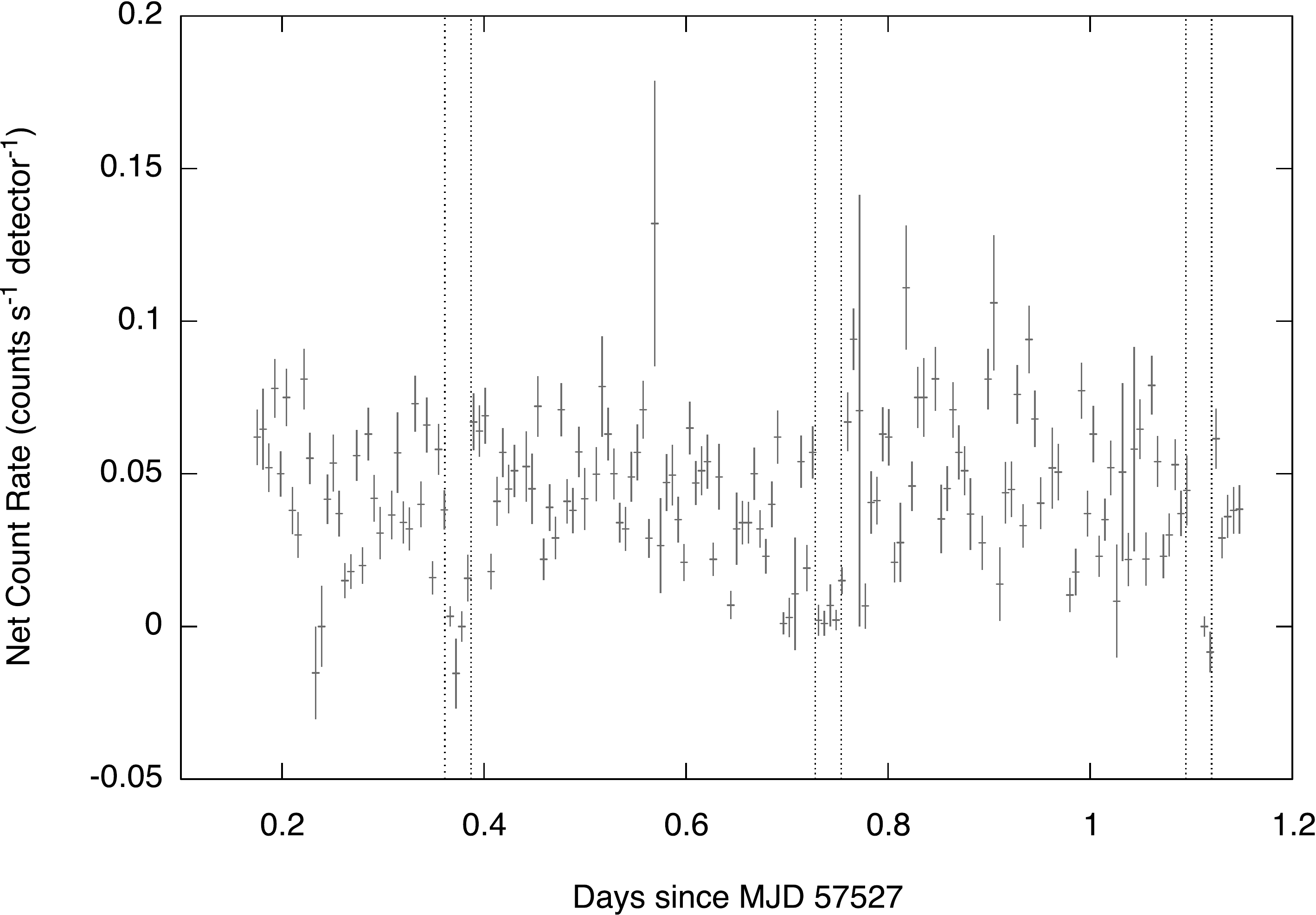}
\caption{The background-subtracted \textit{NuSTAR} light curve (average of FPMA and FPMB; 3--79~keV) of 3FGL J0427.9--6704 binned in 500s intervals. The three X-ray eclipses are indicated by vertical dashed lines.}
\label{fig:nu_lc}
\end{figure*}

The \emph{NuSTAR} light curve, showing effective count rate in 500 s bins, is shown in Figure \ref{fig:nu_lc}. There is strong factor of a few variability on timescales of hundreds of seconds, as well as intervals where the count rate is consistent with zero. The \emph{NuSTAR} data span about 2.7 cycles of the known 8.8-hr orbital period, though only about 71\% of this full time span is on source. We folded the light curve on the optical ephemeris, using $P_\mathrm{orb}=31684.61$s and setting $\phi = 0$ at a BJD of 2457527.59971. A 30-bin phased light curve is shown in Figure \ref{fig:phase_lc}. At $\phi = 0.75$ we detect an X-ray dip with a mean count rate of $(3.4\pm2.1)\times10^{-3}$ cts s$^{-1}$, compared to a mean count rate of $5.4\times10^{-2}$ cts s$^{-1}$ ($\sigma = 1.2\times10^{-2}$ cts s$^{-1}$) for the off-dip phases. Given that the X-ray flux formally drops by a factor of $\sim 16$ and is consistent with zero, and that three separate dips are observed at the same phase (Figure \ref{fig:nu_lc}), we identify these events as X-ray eclipses. 

To better constrain the length of the eclipse, we used a finer binning, finding that 500 bins (each of about 63.4 sec and on average about 10 counts per bin) provided a good balance between time resolution and per bin statistics. We found that the X-ray eclipse has a duration of about $2200\pm60$ sec. The depth and duration of each individual eclipse are consistent with the value determined from the phased light curve. As we detected only about 20 counts during eclipse total, the spectral information is not well constrained. For the same reason, the non-eclipse X-ray spectrum is essentially identical to the average spectrum presented earlier, with the flux scaled up by 6.0\% to correct for the source ``off" time.

In addition to the eclipse, there is some evidence for low-frequency modulation in the phased light curve, with a maximum around $\phi\sim0.1$ and a minimum somewhere in the range $\phi\sim0.5$--0.7. To estimate the significance of this variation, we compared the average count rates between $\phi =$ 0--0.5 and $\phi =$ 0.5--1.0 (excluding the eclipse). The difference in count rates is $(1.02\pm0.17)\times10^{-2}$ cts s$^{-1}$, suggesting evidence for orbital variability (in addition to the eclipse) at over $5\sigma$. Consistent with the spectral analysis presented above, these variations could be due either to changes in $N_H$ over the orbit or with a real change in the X-ray spectrum.

Finally, as a first test for the presence of X-ray pulsations, we divided the \emph{NuSTAR} data into 10 segments and searched for periodic signals using \texttt{efsearch}, with time steps of 10 msec for periods $> 100$ msec and time steps of 0.1 msec for shorter periods. No significant signals were found. However, we emphasize that such searches are extremely challenging and rarely successful unless the radio ephemerides are known. As an example, the X-ray pulsations in the XMM observations of XSS J12270--4859 were not found in an initial search by de Martino et al.~(2013) but were discovered once the radio ephemeris were known (Papitto et al.~2015).

\subsubsection{Fermi Pass 8}

We analysed the PASS 8 \emph{Fermi}/LAT data of 3FGL J0427.9--6704 during the first 7.5 years since the mission (from 2008 August 04 to 2016 February 20; $\mathrm{ROI}=20^\circ$) to try to improve the source localization and to better constrain the light curve. Using the \texttt{Fermi Science Tools} (v10r0p5), we performed a binned analysis with a source model based on the 3FGL catalog (Acero et al.~2015). All catalogued sources within 30$^\circ$ of our target were included. Spectral parameters of the field sources within $7^\circ$ from the target and a few variable sources outside $7^{\circ}$ were allowed to vary during fitting (for sources at distances of $6^\circ-7^\circ$ and the distance variable sources, only the normalizations were freed).  In the course of this analysis, we detected an additional ``source" with strong $\gamma$-ray emission (Test Statistic (TS) = 188, equivalent to a detection significance of $13.7\sigma$) $1.6^\circ$ away from the target, far in the outskirts of the LMC. This could be a true uncatalogued source or residual LMC emission. In any case, an extra point source with a power-law spectrum was added to minimize this source of unknown emission.

Using a binned likelihood analysis, we fit the  \emph{Fermi}/LAT spectrum of 3FGL J0427.9--6704 spectrum with a power law. The best-fit spectral parameters are $\Gamma_{\gamma}=2.48\pm0.06$ and $F_\mathrm{0.1-300GeV}=(9.4\pm0.6)\times10^{-12}$\flux\ with $\mathrm{TS}=441$. These parameters are entirely consistent with that in the 3FGL, which are based on the Pass 7 reprocessed data of the first 4 years ($\Gamma_{\gamma}=2.46\pm0.07$ and $F_\mathrm{0.1-100GeV}=(9.4\pm0.8)\times10^{-12}$\flux). This is good evidence that the source did not vary in flux or spectrum over the last $\sim 4$ years. The spectral fit is plotted in Figure \ref{fig:pow}.

There is no evidence for curvature in the spectrum, as would be expected for an exponential cutoff: if this is left as a free parameter in the fit, it defaults to the high limit of the range allowed, and the quality of the fit decreases. The spectrum of 3FGL J0427.9--6704 appears to be a pure power law.

We also ran \texttt{gtfindsrc} to optimize the source location based on the best-fit source model. The optimized coordinates are J2000 right ascension and declination of (04:27:46.23, --67:03:29.6; error radius at 95\% confidence of 2\arcmin). This positional uncertainty is much improved from 3FGL, and the X-ray and optical sources still sit well within the error circle, providing yet stronger evidence that the $\gamma$-ray source is associated with the emission at other wavelengths (Figure \ref{fig:dss}).

Using the 0.1--300 GeV events extracted with a $1^\circ$ radius aperture and \texttt{TEMPO2} (Hobbs et al.~2006) with the \texttt{Fermi} plugin, we made a 10-bin $\gamma$-ray phase-folded light curve with phases identical to the X-ray one. This light curve has 4800 counts. The light curve is generally flat with orbital phase with a mean count of $480\pm22$ cts bin$^{-1}$. However, we also observe a dip to $429\pm21$ cts bin$^{-1}$ at $\phi=0.7-0.8$, matching the phase of the X-ray eclipse (Figure \ref{fig:phase_lc}). To check this finding, we performed  likelihood analyses on the LAT data in the eclipse ($\phi=0.713-0.783$) and off-eclipse phases. The best-fit $\gamma$-ray fluxes are $F_\mathrm{0.1-300GeV}=(4\pm2)\times10^{-12}$\flux\ and $F_\mathrm{0.1-300GeV}=(9.8\pm0.6)\times10^{-12}$\flux\ in and out of eclipse, which formally differ at the $2.8\sigma$ level. Thus we view the evidence for a $\gamma$-ray eclipse as suggestive but not yet definitive.

\begin{figure}[t]
\centering
\includegraphics[width=3.4in]{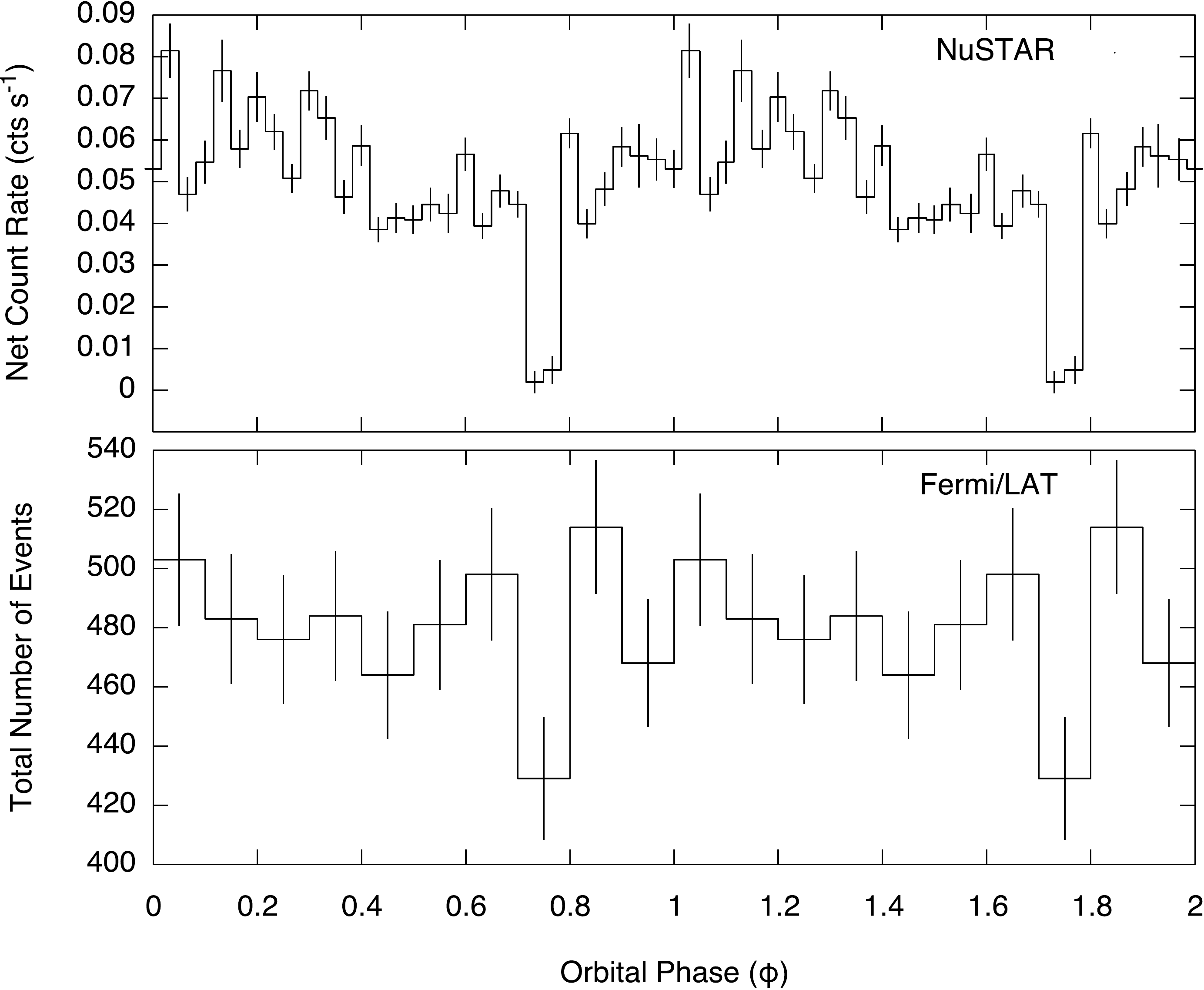}
\caption{The upper panel shows the phase-folded light curve with \textit{NuSTAR} data (3--79~keV; background subtracted) while the lower panel shows the same curve with \textit{Fermi}/LAT data (100~MeV--300~GeV; background not subtracted). The eclipse at $\phi = 0.75$ visible in the \emph{NuSTAR} data also appears to be present in the \emph{Fermi}-LAT light curve, with a formal significance of $2.8\sigma$.}
\label{fig:phase_lc}
\end{figure}

\subsection{X-ray Eclipse and Dynamical Constraints}

The duration of the \emph{NuSTAR} X-ray eclipse outlines a specific curve in the $q$--$i$ plane between mass ratio and inclination, subject only to the assumption that the X-rays are emitted from a central point source and the donor is filling its Roche lobe (Chanan et al.~1976). The X-ray eclipse sets a lower limit on the mass ratio of $q \gtrsim 0.12$ (for $i=90^{\circ}$). Total X-ray eclipses in low-mass X-ray binaries are typically observed to arise for inclinations $\sim 75$--80$^{\circ}$, with inclinations $\gtrsim 80^{\circ}$ associated with only partial eclipses or accretion disk corona sources, likely due to obscuration of the central source by vertical structure in the disk (e.g. Frank et al.~1987). This 
75--80$^{\circ}$ inclination range corresponds to $q \sim 0.28$--0.58. If we take a more generous range of $i$ = 75--85$^{\circ}$, the lower limit decreases to  $q \sim 0.15$.

\begin{figure}[t]
\centering
\includegraphics[width=3.4in]{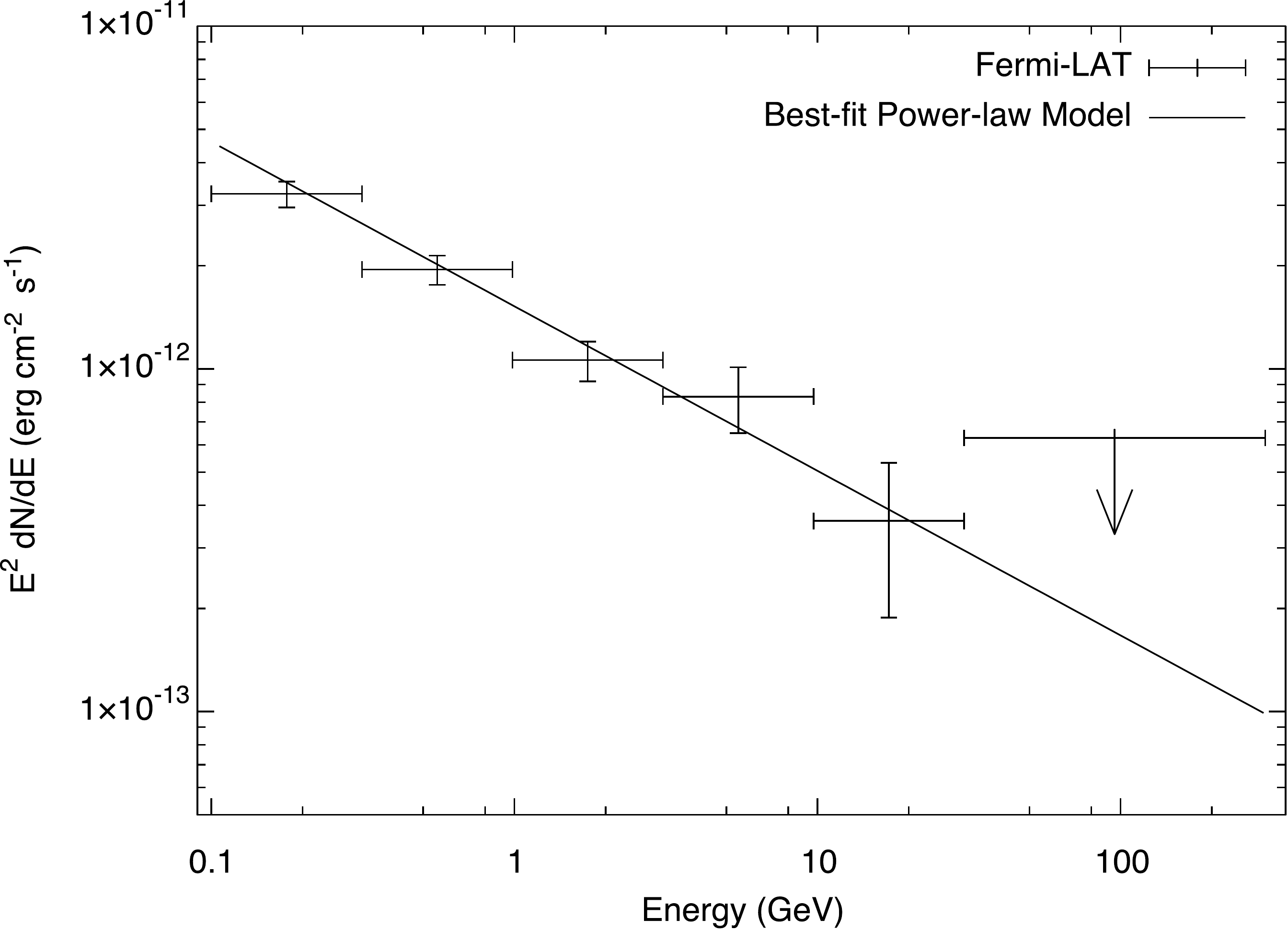}
\caption{\emph{Fermi}-LAT spectrum of  3FGL J0427.9--6704 with the best-fit power law overplotted. The highest energy bin is a 95\% upper limit.}
\label{fig:pow}
\end{figure}

We have an important additional constraint: the $K_2$ measurement of $293\pm4$ km s$^{-1}$ implies a mass function $f(M) = PK_2^3/2\pi G = 0.96\pm0.04 M_{\odot}$ for period $P$ and gravitational constant $G$. Given that $f(M) = M_1 \, {\rm sin}^{3} i/(1+q)^2$, we still require one additional measurement to fully determine the system parameters. If we interpret our measurement of $K_{em} = 103\pm8$ as $K_1$, then we have $q = 0.35\pm0.03$ and $i = 78.3\pm0.6^{\circ}$, and the component masses are $M_1 = 1.86^{+0.11}_{-0.10} M_{\odot}$ and $M_2 = 0.65\pm0.08 M_{\odot}$. This analysis has important caveats in the assumption that $K_{em} = K_1$ and that the measurement of $K_2$ is unaffected by irradiation, but nonetheless is our best dynamical estimate of the masses. 

Another approach is to consider the relations of Casares (2015; 2016), who showed empirically that the mean shape of the H$\alpha$ emission profiles for accreting stellar-mass black holes and white dwarfs are strongly correlated with dynamical properties of the system, including $K_2$ and $q$. To use these relations, we constructed a mean H$\alpha$ profile from the 22 spectra covering that region, then fit both single and double-Gaussian models. We find that the median H$\alpha$ full-width at half-maximum for single-Gaussian fits is $1502\pm41$ km s$^{-1}$. Using the black hole relation, appropriate for low values of $q$, the implied orbital semi-amplitude of the secondary is $K_2 = 350$ km s$^{-1}$. This relation depends weakly on $q$, and for the less extreme mass ratios expected for 3FGL J0427.9--6704, the results of Casares (2015) suggest that a lower value of $K_2 \sim 285$--300 km s$^{-1}$ would be favored. This range is in remarkable agreement with our measured value of $K_2$. 

Following Casares (2016), a constraint on $q$ arises from the ratio of the peak separations to the full-width at half-maximum of H$\alpha$, which is $0.561\pm0.004$. There is a similar systematic issue to the estimation of $K_2$ from the H$\alpha$ full-width at half-maximum, which is that we are in a regime where the exact relations derived by Casares (2016) for black holes and white dwarfs do not precisely apply, giving a range of $q \sim 0.09$--0.35 depending on the assumptions made. This range, while broad, is generally in agreement with the values inferred from the arguments above.

\subsection{Optical and Near-IR Light Curve Fitting}

The preceding sections provide evidence that 3FGL J0427.9--6704 is an eclipsing low-mass X-ray binary and outline some basic constraints on its properties. Here we investigate the additional constraints possible using the optical and near-IR light curves: in particular, the length and depth of the primary and secondary eclipses in the photometry as well as the emission from the irradiated secondary.

We modeled the $BVIH$ light curves, including both the SMARTS and OGLE photometry, using ELC (Orosz \& Hauschildt 2000). The basic model setup was a Roche lobe-filling secondary and a neutron star primary surrounded by an accretion disk, all irradiated by an X-ray source. We assumed a circular orbit. The free parameters were the inclination $i$, the mass ratio $q$, the inner and outer radii of the disk ($r_{in}$ and $r_{out}$), the inner disk temperature $T_d$ and the power-law index of the disk temperature profile $\xi$, the opening angle of the disk rim $\beta$, the mean temperature of the secondary $T_2$ and its mass $M_2$ and radius $R_2$, and the central X-ray luminosity $L_X$. The two external constraints placed on the fits were that they match the length of the X-ray eclipse and the measured $K_2$ value of the secondary. We note that fits of essentially identical quality can be obtained for a range of neutron star masses, and that the value of $K_2$ is by far the most important constraint on the mass in the fit.

Before fitting the photometry, we corrected the magnitudes assuming a foreground reddening of $E(B-V) = 0.037$ (Schlafely \& Finkbeiner 2011). However, there is also the possibility that there is addition extinction local to the source: the $N_H$ inferred from the X-ray spectral fitting is extremely high ($5.8 \times 10^{22}$ cm$^{-2}$). The many magnitudes of optical extinction implied by this value are clearly ruled out by the UV/blue emission from this system; hence we believe the ``excess" $N_H$ producing photoelectric absorption is not associated with dust. We did attempt fits allowing the extinction to vary but they were not significant improvements, so we only assume foreground extinction for our analysis. We cannot rule out the possibility that there is an additional component of extinction that selectively affects only certain portions of the emitted light, but such modeling is much more complex and beyond the scope of this initial paper. 

Given that a perfect fit was not obtained, and that the scatter in the light curve at each phase is much larger than the photometric uncertainties, it would be misleading to report exact uncertainties for each derived parameter. Instead, we discuss the best fit that appears to credibly represent most of the observed features of the system.

The best fit is shown in Figure \ref{fig:mod_lc}, with the parameters listed in Table \ref{tab:dat2}. This fit has a neutron star mass of 1.79 $M_{\odot}$. Qualitatively, this model fits all the relevant aspects of the system, with one exception: the depth of the $I$-band primary eclipse is about 0.1 mag too shallow. However, the eclipses in $B$, $V$, and $H$ are all well-matched. We found no combination of model parameters that could  better match the total set of observations. Non-standard extinction could affect these fits, though it would be expected to affect the bluer bands more.

\begin{deluxetable}{lr}
\tablecaption{Parameters for Light Curve Fitting \label{tab:dat2}}
\tablehead{Parameter    &  Value  }
\startdata
$q$ & 0.335 \\
$i$ ($^{\circ}$) & 79.9 \\
$r_{in}$ ($R_{\odot}$) & $5.4\times10^{-5}$  \\
$r_{out}$ ($R_{\odot}$) & 0.98 \\
$T_{d}$ (K) & $3.57\times10^{5}$\\
$\xi$ & --0.423 \\
$\beta$ ($^{\circ}$) & 0.8 \\
$T_2$ (K) & 4290 \\
$M_1$  ($M_{\odot}$)& 1.79  \\
$M_2$  ($M_{\odot}$)& 0.60  \\
$R_2$ ($R_{\odot}$) & 0.83 \\
$L_X$ (erg s$^{-1}$)& $2.7\times10^{35}$  \\
$d$ (kpc) & 2.4 \\
\enddata
\label{tab:lc_tab}
\end{deluxetable}

Consistent with the discussion in \S 3.4, the region of best fits has a single independent parameter which could be equally well taken as $M_1$, $q$ or $i$. Lower inclinations imply a more massive secondary and thus a physically larger system, increasing the best-fit outer radius of the disk and the effective radius of the secondary. For the absolute best fitting model, which has $M_1 = 1.79 M_{\odot}$, $q=0.335$ (implying $M_2 = 0.6 M_{\odot}$), and the corresponding best-fit inclination is 79.8$^{\circ}$. The mapping between $q$ and $i$ is slightly different in these ELC models than assumed in \S 3.4 due to proper modeling of the non-spherical secondary, but the effect is small: this inclination value is about $1.1^{\circ}$ higher at the same value of $q$, irrelevant to any dynamical calculations. We note that the best-fit light curve model is very similar to that inferred from the dynamical modeling in \S 3.4.

In the light curve fitting, secondary masses in the range $\sim 0.3$--0.7 $M_{\odot}$  produce fits of slightly lower quality but which still appear reasonable overall. A possible argument against secondaries in the mass range 0.3--0.4 $M_{\odot}$ is that the best-fit inclinations are 82--83$^{\circ}$, outside the $\sim$ 75--80$^{\circ}$ inclination range discussed above in which total eclipses are typically observed. Nonetheless, nothing in the light curves themselves exclude these secondary masses.

\begin{figure*}[t]
\includegraphics[width=7in]{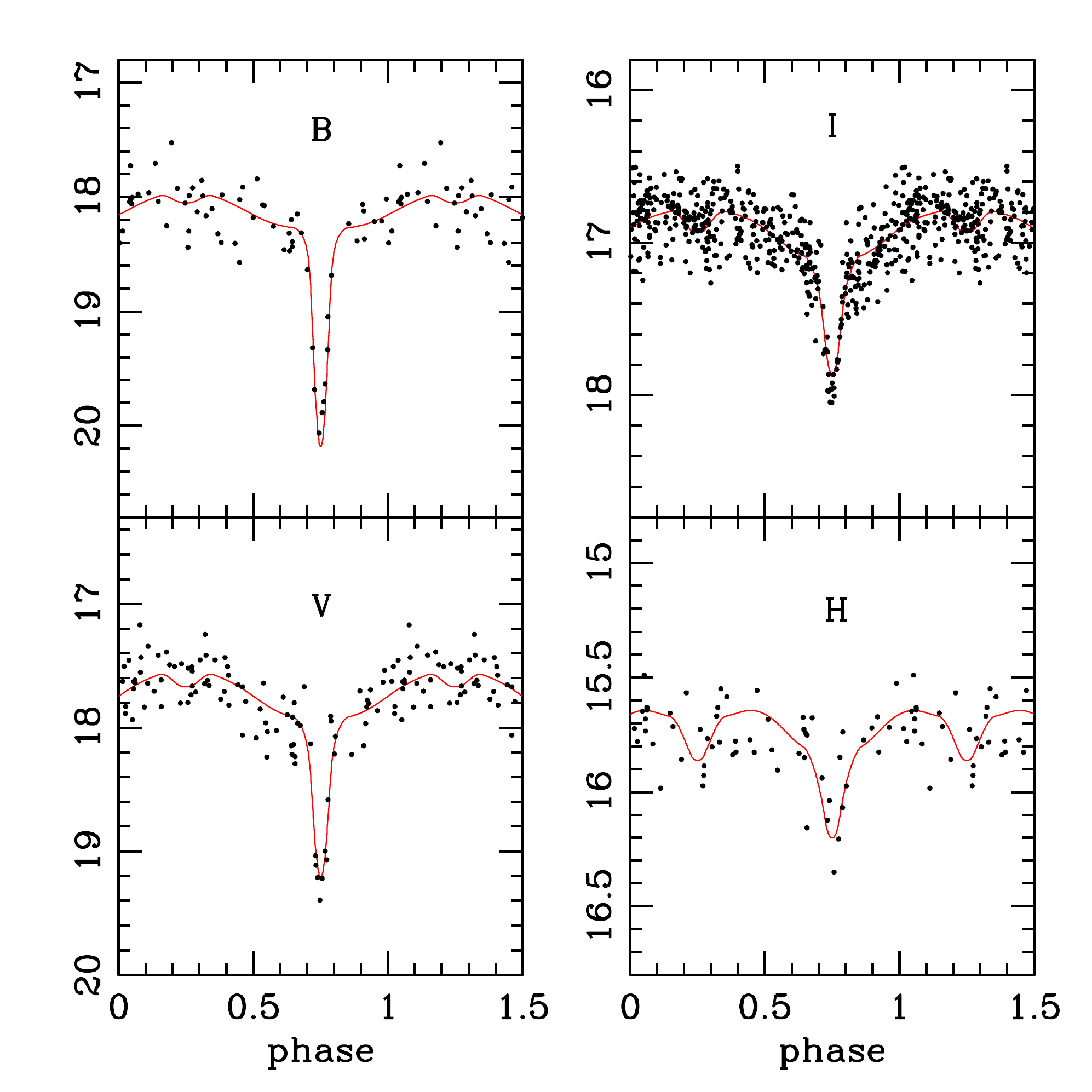}
\caption{ELC model light curve fits to $BVIH$ photometry for the parameters listed in Table \ref{tab:dat2}.}
\label{fig:mod_lc}
\end{figure*}

The ELC model has a large number of parameters to define the disk. The best-fit inner disk radius is 38 km; while reasonable (e.g., Cackett et al.~2009), this value is strongly degenerate with other parameters in the fit and not well-constrained. The outer disk radius is $0.98 R_{\odot}$ and is better constrained by the optical eclipses (up to the physical scale of the binary, i.e., lower component masses would yield a smaller radius). This outer radius is, as expected, much larger than the circularization radius of $0.52 R_{\odot}$ for this system, and is about 0.72 of the primary Roche lobe radius $R_1$. The outer radii of accretion disks in low-mass X-ray binaries are generally argued to be set by the disk tidal radius, which is typically assumed to be 0.9 of $R_1$ (Frank et al.~2002), so we find an outer radius about 20\% lower than this value. The disk temperature profile $\xi$ is remarkably close to the value of --3/7 theoretically predicted for an irradiated disk (Hayakawa 1981) and is much shallower than the --3/4 index for a standard Shakura-Sunyaev disk. The rim opening angle is smaller than inferred for many X-ray binaries (e.g., de Jong et al.~1996) but again is probably not well-constrained. Other than the outer radius, none of the other disk parameters vary substantially with the secondary mass.

A byproduct of the model is the relative flux of the disk and star at each band, which generally varies throughout the orbit as the irradiated side of the star rotates in and out of view. Excluding the primary eclipse, the mean ratio of disk to stellar light in $BVIH$ is $\sim 3.2$, 1.7, 1.0, and 0.5 respectively. While we do not explicitly use the \emph{Swift}/UVOT UV photometry in our fits, considering the bands expected to be dominated by the disk ($V$ and blueward), we note that a single power law fits the disk component of $B$ and $V$ and all of $uvw2$, $uvm2$, and $uvw1$ extremely well: $F_{\nu} \sim \nu^{-1.4}$ with a scatter of only 0.05 dex in log $F_{\nu}$. This spectral behavior, with flux density decreasing toward the UV, is as expected for an irradiated disk (e.g., Frank et al.~2002).

The X-ray luminosity deserves special discussion. This quantity enters the model only inasmuch as it serves to set the observed temperatures of the irradiated star and disk. In the version ELC we used, the X-ray source is modeled as a thin disk surrounding the compact object, which for our system may be an inappropriate physical model (if, for instance, the X-ray source is instead located out of the disk, as in a corona). The thin disk will be foreshortened to surface elements on the disk and the unshadowed elements toward the stellar equator in a manner different than for a extended source located above/below the compact object. The $\gamma$-rays may also be important in irradiating the disk and secondary. Therefore we do not interpret the model X-ray luminosity as the likely true luminosity and in fact, as we argue below, the real luminosity is likely to be much smaller.

Overall, the light curves offer important constraints on the properties of the disk and secondary but do not substantially improve the inferences on $q$ or $i$. This is perhaps unsurprising, as the emission from the secondary is dominated by irradiation rather than ellipsoidal modulations. It is heartening that the overall best-fit (with $M_1 = 1.79 M_{\odot}$; $q = 0.335$, and $i=79.8^{\circ}$) implies $K_1 = 98$ km s$^{-1}$, in excellent agreement with the values inferred from the emission line and disk absorption line radial velocity curves,  keeping in mind that those values may not be faithful measurements of $K_1$. 

As noted above, reasonable fits to both the light curves and the measured value of $K_2$ can be found for a modest range of $q$. More extreme mass ratios would produce lower primary and secondary masses, but also higher inclinations $> 80^{\circ}$ less often observed among eclipsing low-mass X-ray binaries, and values of $K_1$ less consistent with the observed value of $K_{em}$. For example, a $0.5 M_{\odot}$ secondary would have $i=80.7^{\circ}$ and $K_1 = 87$ km s$^{-1}$. The opposite conclusions would hold for more massive secondaries, and indeed we cannot entirely exclude very massive ($> 2 M_{\odot}$) primaries.

In general, we conclude that our best-fit model represents all observed aspects of the systems well but that there is still some latitude in the precise values of the component masses. Our light curve model predicts that the secondary makes a dominant contribution in the near-IR, hence, spectroscopy in this region could offer a chance for improved dynamical constraints on the system. In particular, the measurement of the rotational broadening of the secondary would allow an independent estimate of $q$.

\subsubsection{Distance}

The system distance implied by the best-fit light curve is 2.4 kpc. This value would change by about 10\% for the modest variations in secondary mass discussed above, but the systemic uncertainties, such as the metallicity of the system, are much larger. At this distance, the binary would be located about 1.5 kpc below the Galactic Plane. This could be consistent with either thick disk membership or with a system that was kicked out of the thin disk.

A separate constraint on the system distance comes from the well-known correlation between a combination of the X-ray luminosity and orbital separation and the optical spectral luminosity density (van Paradijs \& McClintock 1994), which is generally interpreted as evidence that the optical luminosity is due to re-processed X-ray emission. Using the more complete compilation of neutron star low-mass X-ray binaries in Russell et al.~(2007), we find a relatively weak constraint: any distance in the range $\sim 0.4$--6 kpc would be consistent with the fairly large scatter in the correlation. Nonetheless, this constraint is consistent with the more precise values obtained from the light curve fitting.

\subsection{Interpreting the Binary and its High-Energy Emission}

\subsubsection{X-ray Emission and Transitional Millisecond Pulsars}

The signpost of transitional millisecond pulsars in the disk state is X-ray ``mode switching": rapid, factor of $\sim 5$ variations in the X-ray luminosity sometimes associated with a slightly softer spectrum in the high state (Linares 2014). This behavior has been observed in all three of the confirmed transitional systems (de Martino et al.~2013; Linares et al.~2014; Patruno et al.~2014; Bogdanov et al.~2015) as well as in 1RXS J154439.4--112820, which shows comparable X-ray variability (Bogdanov \& Halpern 2015; Bogdanov 2015). In 3FGL J0427.9--6704 we observe rapid X-ray variability, including bins with a count rate consistent with zero outside of eclipse, as well as evidence for a softer spectrum at higher count rates. The clinching evidence for mode switching would be the clear identification of a bimodal count rate distribution, but this is harder to verify for this system than for the others due to our lower count rate and  thus need for coarser binning.

The X-ray spectrum and inferred luminosity provide additional evidence that 3FGL J0427.9--6704 is a transitional system. The \emph{NuSTAR} 3--79 keV eclipse-corrected flux is ($7.1\pm0.4$) $\times 10^{-12}$ erg s$^{-1}$ cm$^{-2}$, giving a luminosity of ($4.9\pm0.3) \times 10^{33} (d/2.4 \, {\rm kpc})^2$ erg s$^{-1}$. To compare to the X-ray properties of the known transitional millisecond pulsars, this luminosity implies 0.5--10 keV luminosity of ($2.4\pm0.2) \times 10^{33} (d/2.4 \, {\rm kpc})^2$ erg s$^{-1}$. In the context of interpreting 3FGL J0427.9--6704 as a candidate transitional millisecond pulsar, we take this luminosity as an average between the high and low modes in the disk state. Given that the known transitional systems have high state disk luminosities of 3--$5 \times 10^{33}$ erg s$^{-1}$ (0.5--10 keV) and spend $\sim$ 60--80\% of the time in the high state (Linares 2014), our observed luminosity of $2.4 \times 10^{33}$ erg s$^{-1}$ is precisely as expected for a transitional millisecond pulsar in the disk state. The photon index observed ($\Gamma \sim 1.7\pm0.1$) is consistent with that of M28I, PSR J1023+0038, and 1RXS J154439.4--112820 in their disk states (Linares et al.~2014; Tendulkar et al.~2014; Bogdanov 2015), and the evidence for a softer spectrum at higher count rates is as observed for XSS J12270--4859 (de Martino et al.~2013; Linares 2014). The 8.8-hr orbital period and low-mass main sequence companion are consistent with the binary properties of the known transitional millisecond pulsars (with orbital periods in the range 4.8--11.0 hr; Archibald et al.~2009; McConnell et al.~2015; Papitto et al.~2013; Bassa et al.~2014; de Martino et al.~2014). 

One possible difference is that 3FGL J0427.9--6704 shows evidence for low-frequency orbital variability in X-rays (\S 3.3.2), while for the known transitional systems, it is only observed in the pulsar state and not for the disk state (Tendulkar et al.~2014; Li et al.~2014; Bogdanov et al.~2015; Bogdanov 2015). This tension can be alleviated if we attribute the orbital variations in 3FGL J0427.9--6704 to variations in $N_H$ due to the high inclination. This is consistent with our observations.

\subsubsection{$\gamma$-ray Emission}

$\gamma$-ray emission is emerging as an important aspect to the phenomenology of transitional millisecond pulsars. In the 2013 transition of PSR J1023+0038 from the pulsar to disk state, the $\gamma$-ray emission increased by a factor of 6 (Stappers et al.~2014; Takata et al.~2014; Deller et al.~2015), and in XSS J12270--4859 the $\gamma$-ray emission decreased by a factor of 2 as it transitioned out of the disk state (Johnson et al.~2015). No variations have been detected for M28I, but this is non-informative due to its location among other $\gamma$-ray source in a globular cluster and larger distance. 

Excluding M28I, the ratio of 0.1-300 GeV $\gamma$-ray luminosity to 0.5--10 keV X-ray luminosity for the three other systems lie in the range 2--4 (Linares 2014; Johnson et al.~2015; Bogdanov \& Halpern 2015; Deller et al.~2015). From our \emph{Fermi} analysis of 3FGL J0427.9--6704, the 0.1--300 GeV flux of the source is ($9.4\pm0.8$) $\times 10^{-12}$ erg s$^{-1}$ cm$^{-2}$, equivalent to a luminosity of ($6.5\pm0.6) \times 10^{33} (d/2.4 \, {\rm kpc})^2$ erg s$^{-1}$. This gives a $\gamma$-ray to X-ray ratio of $\sim 2.7$, right in line with the other transitional millisecond pulsars. 

There are possibly interesting differences among the $\gamma$-ray spectra of these systems. For both XSS J12270--4859 and PSR J1023+0038 the $\gamma$-ray spectrum appears best fit by a power-law with an exponential cutoff in the disk state (Takata et al.~2014; Johnson et al.~2015); for 3FGL J1544.6--1125 the 3FGL catalog reports a power-law fit with $\Gamma = 2.36\pm0.08$ (Acero et al.~2015). We find that the \emph{Fermi} spectrum of 3FGL J0427.9--6704 is well-fit by a power-law with $\Gamma = 2.48\pm0.06$ and that a single power-law is strongly preferred over a model with an exponential cutoff. 

We note that the $\gamma$-ray spectrum of 3FGL J0427.9--6704 is dissimilar to that of pulsars detected with \emph{Fermi}-LAT, which tend to have a shallower ($\Gamma < 2$) power law-like spectrum at lower energies and a sub-exponential cutoff at a few GeV. Hard power-law spectra are observed for pulsar wind nebulae, in which the relativistic wind of a young pulsar interacts with surrounding material, producing high-energy photons through synchrotron and inverse Compton emission (Abdo et al.~2013).

This simple power-law spectrum of 3FGL J0427.9--6704 is reminiscent of the ``huntsman"\footnote{Millisecond pulsar binaries in which the pulsar is ablating a low-mass companion are named black widows or redbacks, depending on the companion mass (Roberts 2013). As as pulsar in 1FGL J1417.7--4407 has a giant companion in a much longer orbit than any known black widows or redbacks, in keeping with the theme of spiders, we have suggested the term ``huntsman" as being appropriate: this is a family of large spiders that do not engage in sexual cannibalism.} system 1FGL J1417.7--4407, which has a comparable $\gamma$-ray photon index with no evidence for curvature (Strader et al.~2015). This system shows both similarities and differences to the known transitional systems in the disk states: it has a hard power-law X-ray spectrum and double-peaked H$\alpha$ that suggests an accretion disk, but also has a giant companion in a longer period rather than a main sequence companion, a higher ratio of $\gamma$-ray to X-ray luminosity ($\sim 20$), and was visible as a radio pulsar at similar times to the optical observations (Camilo et al.~2016). One interpretation of 1FGL J1417.7--4407 is that despite the presence of a disk, all of the accretion flow is being ejected in a propeller and thus none reaches the neutron star surface to halt the pulsar mechanism.

The other notable feature of the $\gamma$-ray emission from 3FGL J0427.9--6704 is the tentative evidence for a $\gamma$-ray eclipse. Taken at face value, our measurements suggest a partial eclipse, indicating the $\gamma$-ray emission comes from a large region that is not entirely blocked by the secondary during the X-ray eclipse. At present the evidence for an eclipse is just under $3\sigma$. The measured flux during the eclipse is only $2\sigma$ above zero, and so the possibility of a total eclipse, which would imply a smaller $\gamma$-ray production region, cannot be excluded. We note that Romani et al.~(2015) have found tentative evidence for the eclipse of the magnetospheric $\gamma$-ray emission in the non-accreting millisecond pulsar binary PSR J2215+5135.

As many of recent papers cited above have extensive discussions of the physical models for transitional millisecond pulsars, a detailed recapitulation here is superfluous. Considering the main scenarios for $\gamma$-ray production in the disk state: one possibility is that the $\gamma$-rays originate either in a strong shock between the pulsar wind and surrounding material associated with the accretion flow (e.g., Stappers et al.~2014) or due to inverse Compton scattering of disk photons off the pulsar wind (e.g., Takata et al.~2014). These scenarios are disfavored by the recent observations of coherent X-ray pulsations in the disk states of PSR J1023+0038 and XSS J12270--4859 (Archibald et al.~2015; Papitto et al.~2015). The material reaching the surface would be expected to quench the pulsar wind. An alternative scenario is synchrotron self-Compton at the interface between the pulsar magnetosphere and the accretion disk (e.g., Papitto et al.~2014; Deller et al.~2015). For this latter model, the $\gamma$-rays would originate in a small region close to the neutron star, so a total (rather than a partial) eclipse would be expected. 

It is clear that the details of the $\gamma$-ray eclipse in 3FGL J0427.9--6704 represent important constraints on the $\gamma$-ray production mechanism that can be improved with  future \emph{Fermi} observations.

By definition, confirmation of a system as a transitional millisecond pulsar can only come via an observed transition between the disk and pulsar states. For 3FGL J0427.9--6704 the OGLE photometry implies the existence of a similar accretion disk since 2010, so the system has been visible as a low-mass X-ray binary for at least six years. The constancy of the \emph{Fermi} flux over a similar timescale (\S 3.3.3) is additional evidence for no transition during this time. Thus, as for the case with 1RXS J154439.4--112820 (which has been in the disk state for at least 10 years; Bogdanov 2015) we are in somewhat of a waiting game to see whether a transition will occur. In any case, the discovery of these systems shows that we are far from complete in our census of $\gamma$-ray bright compact binaries associated with \emph{Fermi} sources.

\acknowledgments

We thank an anonymous referee for helpful comments that improved the paper. We thank C.~C.~Cheung, R.~Corbet, T.~Maccarone, D.~Nidever, T.~Tauris, and L.~van Haaften for useful conversations, F.~Walter for providing routines and advice for the reduction of ANDICAM photometry, and R.~J.~Foley and Y-C.~Pan for data acquisition. This project made use of \emph{NuSTAR} mission data, a project led by the California Institute of Technology and managed by the Jet Propulsion Laboratory. We sincerely thank F.~Harrison, K.~Forster, and the \emph{NuSTAR} team for scheduling and executing the \emph{NuSTAR} observations. Based on observations obtained at the Southern Astrophysical Research (SOAR) telescope, which is a joint project of the Minist\'{e}rio da Ci\^{e}ncia, Tecnologia, e Inova\c{c}\~{a}o (MCTI) da Rep\'{u}blica Federativa do Brasil, the U.S. National Optical Astronomy Observatory (NOAO), the University of North Carolina at Chapel Hill (UNC), and Michigan State University (MSU). The Digitized Sky Surveys were produced at the Space Telescope Science Institute under U.S. Government grant NAG W-2166. This publication makes use of data products from the Wide-field Infrared Survey Explorer, which is a joint project of the University of California, Los Angeles, and the Jet Propulsion Laboratory/California Institute of Technology, and NEOWISE, which is a project of the Jet Propulsion Laboratory/California Institute of Technology. WISE and NEOWISE are funded by the National Aeronautics and Space Administration. We acknowledge the use of public data from the Swift data archive. The OGLE project has received funding from the National Science Centre, Poland, grant MAESTRO 2014/14/A/ST9/00121 to AU. Support from NASA grant NNX15AU83G is gratefully acknowledged.

{}

\end{document}